# Current-induced creation and dynamics of embedded magnetic skyrmion bags

Yaodong Wu[1], Jialiang Jiang[2, 3*], Lingyao Kong[2], Meng Shi[3], Shouguo Wang[4], Mingliang Tian[2, 3], Haifeng Du[3*], and Jin Tang[2, 3*]

[1]School of Physics and Materials Engineering, Hefei Normal University, Hefei 230601, China

[2]State Key Laboratory of Opto-Electronic Information Acquisition and Protection Technology, School of Physics, Anhui University, Hefei, 230601, China

[3]Anhui Provincial Key Laboratory of Low-Energy Quantum Materials and Devices, High Magnetic Field Laboratory, HFIPS, Chinese Academy of Sciences, Hefei 230031, China

[4]Anhui Provincial Key Laboratory of Magnetic Functional Materials and Devices, School of Materials Science and Engineering, Anhui University, Hefei 230601, China

*Corresponding author: jintang@ahu.edu.cn; duhf@hmfl.ac.cn; jjl2024@ahu.edu.cn

## Abstract

Magnetic skyrmion bags—vortex-like structures hosting multiple skyrmions with tunable topological charge ($Q$)—hold significant promise for next-generation spintronic computing. However, while their creation using magnetic fields has been demonstrated, their direct electrical generation remains an outstanding challenge. Here, we report the direct current-induced formation and manipulation of embedded skyrmion bags in a FeGe nanoplate under zero magnetic field. Using in-situ Lorentz transmission electron microscopy, we capture the transformation of a distorted helical ground state into embedded skyrmion bags with diverse configurations, driven by nanosecond current pulses. Theoretical analysis indicates that this process is driven by the spin-transfer-torque-induced fracture of the helical state. Furthermore, we demonstrate electrically-induced transitions between skyrmion bags of different $Q$, leading to the stabilization of complex three-dimensional topological structures, including experimental signatures of magnetic monopoles and bobbers. Our work establishes a foundation for all-electrical control of high-$Q$ topological spin textures and topological defects, paving the way for their application in functional spintronic devices.

## Introductions

Charge quantization of nuclei is the fundamental law of physics, giving rise to the colorful world we live in. In magnetic materials, the magnetization of magnetic structures can be quantified using the invariant known as the topological charge $Q$, which essentially counts the number of times the physical space covers the unit sphere that the magnetization vector **m** points to.[1] Likewise, the quantization of topological charges determines various topological spin textures, such as magnetic skyrmions[2-6], magnetic bobbers[7], and magnetic merons[8]. Topological spin textures with emergent electromagnetic properties are promising information carriers applied in spintronic devices[9-15]. Among these topological spin textures, magnetic skyrmions were the first to be discovered[3], characterized by their unit $Q$, whose value depends on the specific magnetic arrangement configuration, either being $Q = 1$ or $-1$.[4-6,16] Subsequent advancements have employed skyrmions and counterparts as discrete building units to construct topological spin textures with customizable topological charges $Q$[17-22], which may expand the potential applications of topological spintronic devices, including ASCII binary information encoding and interconnect devices[17,23].

In a two-dimensional context, multiple-$Q$ magnetic solitons have been envisioned as skyrmion bags or high-order skyrmions[17,18,24-27]. We introduce a notation $S(N)$ to represent skyrmion bags having $N$ interior skyrmions with $Q = 1$ and a total degree of $Q = N - 1$. This definition can also be expanded to more complex structures with $S(n_2)$ skyrmion bags inside a $S(n_1)$ skyrmion bag, called nested $S(n_1, S(n_2))$ skyrmion bags with $Q = n_1 - n_2$, as shown in Fig. 1. Advancing into three-dimensional topological magnetism, both theory and experiments confirm the existence of skyrmion bundles[19,21], composed of high-order skyrmions at near-surface layers and skyrmion bags at interior layers. The exterior spiral boundary of these three-dimensional skyrmion bundles is topologically akin to magnetic Hopfions[21,28].

The realization of multiple-$Q$ magnetic solitons poses significant challenges due to the inherent contradictions in the magnetic field that arise during the stabilization of internal skyrmion clusters and boundary spirals. Previous studies have addressed these challenges by employing a two-step process: first, creating internal skyrmion clusters at a negative field, and then reversing the magnetic field to form boundary spirals[19,21,28]. This two-step approach has proven effective in generating diverse multiple-$Q$ magnetic solitons[19,21,28]. Although simulations suggest the creation of multiple-$Q$ solitons using electrical methods[29], experimental confirmation remains elusive. The electrical creation of these solitons represents a crucial step towards the practical application of multiple-$Q$ topological spintronics.

In this study, we experimentally demonstrate the electrical creation of skyrmion bags exhibiting multiple $Q$ values and morphologies in FeGe lamella at zero magnetic fields. This achievement leverages the two-fold degeneracy of skyrmions at zero magnetic fields, where both skyrmions with $Q = -1$ and 1 are stable. This degeneracy provides a natural platform for assembling skyrmion bags. By utilizing current-induced spin-transfer torque[30], we manipulate skyrmions with $Q = 1$ that are encircled by skyrmions with $Q = -1$ (and vice versa), resulting in the formation of diverse nested skyrmion bags embedded in a skyrmion-lattice background magnetization[26]. Furthermore, we report the observation of the current-induced collapse of skyrmion tubes within these nested skyrmion bags. This collapse process leads to the evidence of the emerging topological defects, such as monopoles and bobbers[7,31], which are controlled by the applied currents. These findings not only demonstrate the feasibility of electrically creating multiple-$Q$ magnetic solitons but also open new avenues for exploring and manipulating topological defects in magnetic systems.

## Results

## Electrical creation of magnetic skyrmion bags

A 100-nm-thick FeGe lamella[32], equipped with two electrodes, was fabricated for *in-situ* Lorentz transmission electron microscopy (TEM) imaging[33,34], as shown in Fig. 1a. In the $B_{20}$ FeGe magnets, the skyrmion size is about 70 nm [19,32]. Herein, we demonstrate the creation of magnetic skyrmion bags utilizing pulsed currents with a duration of 70 ns. Notably, the outermost magnetizations of both $S(n)$ and $S(n_1, S(n_2))$ bags exhibit an out-of-plane up orientation, as shown in Fig. 1c. Conversely, when the outermost magnetizations of skyrmion bags exhibit an out-of-plane down orientation, we refer to them as $-S(n)$ bags, with a topological charge of $Q = 1 - n$, and $-S(n_1, S(n_2))$ bags, with a topological charge of $Q = n_2 - n_1$.

Fig. 2 presents a comprehensive overview of the current-driven dynamic transformations, commencing from the initial helix state. The current pulses have a duration of 70 ns. We define a positive current pulse as one in which the current flows from the left to the right side of the device, and a negative current pulse as one in which the current flows from the right to the left side. Notably, when the current density $j$ remains below $3.5 \times 10^{10}$ A/m$^2$, the helix domains exhibit no discernible dynamic responses (Supplementary Fig. 1a). However, as the current density rises within the range of 3.5 to $5.0 \times 10^{10}$ A/m$^2$, we observe a transformation from the helix state to a skyrmion lattice, as depicted in Fig. 2b and 2c, Supplementary Fig. 1b, and Supplementary Movies V1 and V2. Nevertheless, it is crucial to note that once the current density surpasses $5.5 \times 10^{10}$ A/m$^2$, the current-induced helix exhibits chaotic dynamics, primarily attributed to the significant Joule thermal heating effect (Supplementary Fig. 1c and Movie V3). Thus, there present a optimal current density $j_{\mathrm{opt}}$ for current-induced creation of skyrmions and skyrmion bags (Supplementary Fig. 2).

In chiral magnets, the competition between exchange and Dzyaloshinskii-Moriya interaction (DMI) forces the magnetizations to twist periodically on a plane either perpendicularly (helix) or with a canting angle (cone) along a specific direction, defined as $q$-vector [35,36]. For the initial helix (Fig. 2a), there are no preferred $q$-vectors. Once current is applied, the spin-transfer torque forces the formation of a horizontal helix with a $q$-vector perpendicular to the current orientation, as experimentally identified (Fig. 2b). However, we can always observe a perpendicular helix whose top ends are pinned from the geometrical edges. A prior study has conclusively established the creation of skyrmions from the perpendicular helix triggered by spin-transfer torque, which can be attributed to the force analysis of the instability of helix ends[37]. Skyrmions are first split from the perpendicular helix and then move to the bottom side because of skyrmion Hall effects[38,39], resulting in the accumulation of skyrmions. Specifically, for positive currents, we consistently observe the formation of a skyrmion lattice with a negative topological charge, $Q$ (Supplementary Fig. 3). Conversely, reversing the current polarity results in the emergence of a skyrmion lattice with a positive $Q$ (Supplementary Fig. 3).

It is noteworthy that, due to the two-fold degeneracy of skyrmions with $Q = 1$ and $Q = -1$ in the absence of an external field, a few individual skyrmions or closed helix domains with positive and negative $Q$ can both persist (Fig. 2a). As positive pulsed currents are successively applied, these individual skyrmions with positive $Q$ have chances to be embedded into a larger closed helix with negative $Q$ (Fig. 2c). Because of skyrmion Hall effects driven by current, the long helix could undergo the integral crash forces of external skyrmion clusters moving to the bottom ends, contributing to the formation of individual $S(n)$ bags (Fig. 2d). Especially, there has a chance to form $-S(n)$ skyrmion bags embedded into a large closed helix (Fig. 2c). The long helix undergo tension force from the internal bag and can be split to form individual nested skyrmion bags (Fig. 2e). After approximately 120 pulsed currents, the number of skyrmions reaches a stable equilibrium, and no further skyrmion bags are created (Supplementary Movie V1, V2, and V3). Under continuous current stimulation, skyrmions are generated to form a lattice with a uniform topological charge, *e.g.*, $Q = -1$. In contrast, skyrmion bags consist of internal skyrmions with $Q = 1$, which predominantly originate from the initial disordered state (Fig. 2a). As a result, skyrmion bags appear as isolated objects surrounded by a skyrmion lattice, and their number is inherently limited by the availability of oppositely charged seeds.

The temperature rise is closely linked to the current density due to the Joule thermal heating effect[40]. As the current density increases, so does the temperature. Once the temperature increase reaches approximately 250 K, the helix domains become the only stable state at zero field[32]. Therefore, by applying a single high-density current pulse, we can effectively reset the initial helical domains (Supplementary Fig. 4 and Supplementary Movie V4), enabling us to repeat the dynamic process depicted in Fig. 2 and Supplementary Movie V1. In individual experiments, we adopt a standardized procedure. First, we apply a single pulsed current with a density of $5.6 \times 10^{10}$ A/m$^2$ to restore the initial helix domains. Subsequently, we apply approximately 150 current pulses to induce the generation of skyrmions and skyrmion bags.

In our multiple individual experiments, we successfully generated skyrmion bags exhibiting various topological charges, as shown in Fig. 3a. For skyrmion bundles characterized by a depth-modulated spin twisting[19,21], the average in-plane magnetizations of the boundary spirals are significantly weaker compared to those of the interior skyrmions. However, for skyrmion bags, the retrieved average in-plane magnetizations of the boundary spirals are comparable to those of the interior skyrmions. This notable difference can be attributed to the varying background magnetizations. In our present experiments, the skyrmion bag is stabilized within a skyrmion lattice. Consequently, the background magnetization can be approximated as a ferromagnet, leading to negligible spin twists along the depth orientation for skyrmion bags. Our simulations further confirm this observation (Fig. 3b), revealing that the spin textures across all layers maintain their skyrmion bag characteristics (Supplementary Fig. 5).

We successfully generated not only $S(n)$ skyrmion bags but also intricate nested $S(n_1, S(\mathrm{n}_2))$ skyrmion bags by utilizing pulsed currents, as shown in Fig. 3c and Supplementary Fig. 6. The simulation well reproduced the intricate configurations (Fig. 3d). The formation process of these nested skyrmion bags can be outlined in two distinct steps. Initially, an $-S(n_2)$ skyrmion bag surrounded by a helix domain is created (Fig. 1c). Subsequently, the helix domain undergoes a breakage process, resulting in the emergence of an $S(n_1)$ skyrmion bag that encircles the inner $-S(n_2)$ skyrmion bag, thus giving rise to the nested $S(n_1, S(\mathrm{n}_2))$ skyrmion bags configuration.

Current-induced effects also include Joule thermal heating and the Oersted field. By reversing the current orientation, the topological charges of current-induced skyrmions and skyrmion bags are reversed (Supplementary Fig. 7), ruling out Joule heating as the main origin, since it does not depend on current direction. Our multiphysic field simulation shows a maximum temperature rise of about 14 K for a 70-ns current pulse with a density of $4\times10^{10}$ A m$^{-2}$ (Supplementary Fig. 8). This moderate temperature rise may assist the splitting of skyrmions and skyrmion bags from the ends of a long vertical helix, but the topological sign is determined solely by STT. To assess the possible influence of the Oersted field, we performed multiphysics field simulations of the field distribution at the applied current density (Supplementary Fig. 9). The Oersted field is predominantly in-plane, with a maximum magnitude of approximately 2.86 mT, and its orientation differs between the top and bottom surfaces. In our FeGe devices, skyrmion creation typically requires an out-of-plane field on the order of 100 mT. The Oersted field is therefore (i) purely in-plane, (ii) an order of magnitude smaller, and (iii) spatially varying, making it incapable of nucleating skyrmions. Furthermore, symmetry and force analysis yield four distinct STT-driven dynamical behaviors depending on the pinning ends of the vertical helix and the current orientation (Supplementary Fig. 10). Our experiments reproduce all four predicted schemes (Supplementary Fig. 11). Consequently, we conclude that STT is the primary effect determining the topological sign of current-induced skyrmion bags and skyrmions, while Joule heating plays at most an auxiliary role. Since our devices are based on FeGe without adjacent heavy-metal layers, spin–orbit torques are absent, and the current-induced torque is dominated by the in-plane Zhang–Li STT[30]. The creation mechanism of skyrmion bags relies on the coexistence of skyrmions with opposite topological charges at zero magnetic field, and the STT-driven instability of helix ends—both of which are rooted in topology and dynamics rather than in specific material parameters. Therefore, our experimental findings can, in principle, be generalized to other skyrmion hosting systems, including room temperature chiral magnets and multilayers with interfacial DMI[14].

We define a current cycle as the process of transforming helical domains primarily into a skyrmion lattice, driven by the application of current (Supplementary Movie V1 for reference). Fig. 4a presents the total counts of magnetic skyrmion bags obtained from 50 individual current cycles. Each current cycle consists of a single high-current pulse with a density of 5.5 × $10^{10}$ A/m², followed by 150 low-current pulses with a density of 4.5 × $10^{10}$ A/m². Initially, a few skyrmions with topological charges $Q$ = 1 and $Q$ = −1 are present. Upon application of positive current pulses, $Q$ = −1 skyrmions are generated, whereas $Q$ = 1 skyrmions either gradually annihilate or evolve into skyrmion bags (Supplementary Fig. 12). Our findings indicate that $S(1)$ bags, also known as skyrmionium or 2π-vortex, have the highest likelihood of being generated. Notably, the generation possibility of $S(n)$ skyrmion bags decreases as $n$ increases. This trend can be attributed to the instability of high-$Q$ skyrmion bags during current-driven dynamics, as demonstrated in Supplementary Movies V5 and V6. These Movies reveal transitions from $S(3)$ with $Q$ = 2 to $S(2)$ with $Q$ = 1 and $S(6)$ with $Q$ = 5 to $S(5)$ with $Q$ = 4, indicating a reduction in topological charge during the application of current stimuli. Due to the intricate two-step creation process involved in forming nested $S(n_1, S(n_2))$ skyrmion bags, only a few such bags can be observed. The likelihood of their creation aligns with the trend in the total free energy $E$ of skyrmion bags, as illustrated in Fig. 4b. Our simulations reveal that $S(1)$ bags possess the lowest energy, with the total energy increasing as $Q$ increases. Furthermore, nested skyrmion bags generally exhibit relatively high energies. The skyrmion bags stabilize as high-energy metastable phases, originating from initial skyrmion seeds with different signs of $Q$ contrast to that of the skyrmion lattice. To keep the outer magnetization consistency, the initial skyrmions with $Q$ = 1 must be encircled by a skyrmion with $Q$ = −1, resulting in the formation of skyrmion bags.

The reproducibility of current-induced skyrmion bag creation was tested in multiple microdevices with thicknesses ranging from 80 to 150 nm. Consistent current-induced dynamics were observed across all devices (Supplementary Fig. 13–16). Owing to their high-energy metastable nature, thermal fluctuations promote the transformation from skyrmion bags to isolated skyrmions. Accordingly, our experiments show that the probability of skyrmion bag formation per current cycle decreases with increasing temperature (Supplementary Fig. 17b). In contrast, the probability of skyrmion bag formation per current cycle decreases with increasing pulse width (Supplementary Fig. 17a). Under longer current stimuli, there is a higher probability that skyrmion bags will transform into isolated skyrmions, driven by thermal fluctuations and the continuous application of spin-transfer torque. To assess the stability of skyrmion bags, we monitored the magnetic texture under constant conditions (zero applied current, 95 K, and zero external magnetic field) for over 13 hours (Supplementary Fig. 18). Repeated imaging reveals no detectable changes in the morphology, size, or position of the skyrmion bags, demonstrating their long-term stability.

## Current-induced topological transformations assisted by the emergence of topological defects

We have shown the metastable nature of high-$Q$ or nested bags, which possess higher energies compared to $S(1)$ bags. Given this, it is reasonable to expect topological transitions within skyrmion bags during the application of current stimuli. Here, we further explore the dynamics of magnetic skyrmion bags driven by current, as shown in Fig. 5. During the application of pulsed currents, we observe distinct transformations from $S(n)$ to $S(n - 1)$ bags (Supplementary Movies V5 and V6). These transformations are directly linked to the collapse of skyrmion tubes within the bags. The topological transformations from $S(n)$ to $S(n - 1)$ bags are understood by the instability of skyrmion bags due to the current-induced Joule thermal heating. For longer current density and longer current pulse width, the Joule thermal heating effect becomes more remarkable, causing the easy transformations (Supplementary Fig. 19 and 20). If the current orientation is reversed, based on the symmetry analysis and selective polarity rule (Supplementary Fig. 10), the initial skyrmion bags annihilate during the background magnetization transitions between $Q = -1$ skyrmion lattice and $Q = 1$ skyrmion lattice (Supplementary Fig. 7).

The collapse process of skyrmion tubes can be categorized into two distinct styles[41]. Firstly, there is the merger of two skyrmion tubes, facilitated by topological monopoles in Y-shaped skyrmion tubes. Secondly, a skyrmion tube is broken from the interior, assisted by magnetic bobbers. Although these topological defects emerge during field-driven annihilations of skyrmion tubes[31,42,43], experimental observation using electrical methods has remained elusive. Here, we present compelling evidence of these typical collapsed processes in the context of current-induced topological quantized annihilation of nested skyrmion bags. Starting with an $S(0, S(2))$ nested skyrmion bag as the initial state (Fig. 5a-5c), we observe a transition to an $S(0, S(1))$ nested skyrmion bag after the application of a series of current pulses (Fig. 5d-5e). This topological transformation is achieved through the merging of two internal skyrmion tubes, which exhibit a narrow gap between them and contribute to a reduced Fresnel contrast (Fig. 5e and 5f). This phenomenon is consistently observed in our experiments (Fig. 5d). With further application of pulsed currents, we observe a subsequent transformation from the $S(0, S(1))$ state to an $S(1)$ bag (Fig. 5g-5o). This transition is associated with the collapse of a skyrmion tube (Fig. 5j-5l). Notably, in our experiments, we identify an intermediate state characterized by a very weak, dotted-like Fresnel contrast (Fig. 5j). This observation can be attributed to the emergence of a magnetic bobber during the collapse process (Fig. 5k and 5l). The abrupt changes in Fresnel contrast during the collapsed processes of skyrmion tubes in experiments (Fig. 5p) are highly consistent with simulations (Fig. 5q), providing experimental proof for the stabilization of complex 3D topological spin textures.

The emergent intermediate collapsed state of skyrmion tubes driven by current can be widely achieved for nested skyrmion bags (Supplementary Fig. 21 and 22). Typically, we can obtain various styles of internal contrasts for internal topological configurations of nested skyrmion bags (Supplementary Fig. 22), suggesting the stability of bobbers with various penetration depths. Our findings provide experimental evidence for the collapse mechanisms of skyrmion tubes within skyrmion bags and offer insights into the dynamic behavior of these complex magnetic structures under the influence of current stimuli. The distinction between depth-uniform skyrmion tubes and depth-modulated 3D solitons (such as magnetic bobbers or skyrmion bundles) based on Lorentz TEM contrast differences has been established in previous studies [7,16,19,34]. Following this principle, our observed contrast features are consistent with the presence of 3D topological structures, but we refer to them as experimental signatures rather than definitive proof. For direct and unambiguous reconstruction of the full 3D spin configuration, future work should employ tilt-series Lorentz TEM imaging over a wide angular range (e.g., –60° to +60°) combined with tomographic reconstruction [44-46].

Previous studies have established emergent skyrmion lattice dynamics, including rotation driven by spin torques, thermal gradients, and field gradients[47-49]. Typically, such dynamics are inferred from shifting scattering peaks in reciprocal-space neutron scattering experiments[49]. Skyrmion bags—which can be regarded as topological defects within the lattice—offer a unique opportunity to directly image lattice dynamics in real space. In our observations, skyrmion bags undergo counter-clockwise rotation within the lattice, revealing the counter-clockwise rotation of the entire skyrmion lattice (Supplementary Fig. 23 and Movie V3). These findings suggest a pathway for device applications that exploit skyrmion lattice dynamics by detecting the motion of skyrmion bags (Supplementary Fig. 23d). The electrical detection of skyrmion bags with different topological charges $Q$ can be achieved via the topological Hall effect, whose magnitude is proportional to the $Q$ value within the detection region [50-52]. For more efficient and practical read-out, tunnel magnetoresistance effects—integrated into a magnetic tunnel junction geometry—can be used to distinguish different $Q$ values of skyrmion bags (Supplementary Fig. 23d) [53-55]. This provides a pathway for all-electrical writing, manipulation, and reading of skyrmion-bag-based spintronic devices.

## Discussion

In summary, we have experimentally achieved the first instance of the current-induced creation of magnetic skyrmion bags without the application of an external magnetic field. The unique twofold degeneracy exhibited by skyrmions, coupled with the selective polarity arising from the end instability of the helix when a pulsed current is applied, underlies the electrical creation of these skyrmion bags. Our study further uncovers novel current-induced bag dynamics, encompassing topological quantized annihilations. We propose that the intricate collapsed processes involving skyrmion tubes within nested skyrmion bags could pave the way for the realization of monopoles and bobbers through electrical methods. The creation and topological transformation assisted by emerging topological defects, facilitated solely by currents in the absence of an external field, not only provide further insight into skyrmion physics but also open up new avenues for the development of multiple-$Q$ spintronics.

The stochastic nature of skyrmion bag formation in our zero-field electrical approach, while currently limiting deterministic operation, highlights the governing roles of local domain terminations and current direction. Achieving deterministic electrical creation of skyrmion bags will require further exploration, particularly through local manipulation techniques such as laser irradiation and lithographically defined pinning sites.

## Methods

**Fabrications of FeGe microdevice**

A thin FeGe lamella with a thickness of ~80-150 nm for TEM magnetic imaging was fabricated by the lift-out method using a focused-ion beam and scanning electron microscopy dual beam system (Helios Nanolab 600i, FEI) in combination with a gas injection system and a micromanipulator (Omniprobe 200+, Oxford). The results presented in the main text figures are obtained from the same device (Device #1), which has a length of 4 μm, a width of 3 μm, and a thickness of 100 nm.

**TEM measurements**

Magnetic imaging was carried out using a TEM instrument (Talos F200X, FEI) operated at 200 kV at the Fresnel mode. The objective lens is switched off to provide a field-free condition. A single-tilt liquid-nitrogen specimen holder (Model 616.6 cryotransfer holder, Gatan) was used for varying temperature measurements. The pulsed current with a pulse duration of 20-160 ns and a frequency of 1 Hz was provided by a voltage source (AVR-E3-B-PN-AC22, Avtech Electrosystems). All experiments were performed at zero magnetic field.

**Numerical simulations**

Micromagnetic simulations were performed using MuMax3[56]. The total free energy terms are written as: $\varepsilon = \int_{V_s}\{\varepsilon_{\text{ex}} + \varepsilon_{\text{DMI}} + \varepsilon_{\text{dem}}\}\text{d}\boldsymbol{r}$. Here, exchange energy $\varepsilon_{\text{ex}} = A(\partial_x\mathbf{m}^2 + \partial_y\mathbf{m}^2 + \partial_z\mathbf{m}^2)$, DMI energy $\varepsilon_{\text{DMI}} = D_{\text{dmi}}\mathbf{m}\cdot[\nabla\times\mathbf{m}]$, and demagnetization energy $\varepsilon_{\text{dem}} = -\frac{1}{2}M_{\text{s}}\mathbf{B}_{\text{d}}\mathbf{m}$. Here $\mathbf{m} \equiv \mathbf{m}(x, y, z)$ is the normalized units continuous vector field that represents the magnetization $\mathbf{M} \equiv M_{\text{s}}\mathbf{m}(x, y, z)$. *A*, $D_{\text{dmi}}$, and $M_{\text{s}}$ are the exchange interaction, DMI interaction, and saturation magnetization, respectively. $\mathbf{B}_d$ is the demagnetizing field. We set a typical value for saturation magnetization $M_s$ = 384 kA m$^{-1}$ for FeGe[57]. The exchange interaction $A_{ex}$ = 3.25 pJ/m is determined from the fit to the field-dependence of magnetization evolution[57]. DMI interaction $D_{\mathrm{dmi}} = 4\pi A/L_{\mathrm{D}} =$ 0.5834 mJ m$^{-2}$ is obtained from zero-field spin spiral period $L_D$ = 70 nm[57]. We set the cell size as 2 × 2 × 2 nm$^3$. We obtained the equilibrium spin configurations using the conjugate-gradient method.

## Data availability

The data that support the findings of this study have been included in the main text and Supplementary Information. Raw LTEM data (images and movies) are available from the corresponding authors upon reasonable request.

# Fundings

This work was supported by the National Key R&D Program of China, Grant No. 2022YFA1403603 (H.D.); the Scientific Research Innovation Capability Support Project for Young Faculty, Grant No. SRICSPYF-BS2025073 (J.T.); the Natural Science Foundation of China, Grants No. 52325105 (H.D.), 12422403 (J.T.), U24A6001 (J.T.), 12104123 (Y.W.), 12504118 (J.J.), and 12241406 (H.D.); the Anhui Provincial Natural Science Foundation, Grants No. 2308085Y32 (J.T.) 2508085QA029 (J.J.), and 2508085MA018 (Y.W.); the Natural Science Project of Colleges and Universities in Anhui Province, Grants No. 2024AH030046 (Y.W.); Anhui Province Excellent Young Teacher Training Project, Grant No. YQZD2023067 (Y.W.); the 2024 Project of GDRCYY (No. 217, Yaodong Wu); the China Postdoctoral Science Foundation Grant No. 2023M743543 (Y.W.) and 2025M77050 (J.J.); the Natural Science Project of Colleges and Universities in Anhui Province, Grants No. 2024AH030046 (Y.W.) and 2025AHGXZK20140 (Y.W.); CAS Project for Young Scientists in Basic Research, Grant No. YSBR-084 (H.D.); and Systematic Fundamental Research Program Leveraging Major Scientific and Technological Infrastructure, Chinese Academy of Sciences, Grant No. JZHKYPT-2021-08 (H.D.).

## Author contributions

H.D. and J.T. supervised the project, conceived the idea, and designed the experiments. J.T. performed experiments and micromagnetic simulations with the help of Y.W., M.S., L.K., and J.J.. J.T., H.D., J.J., and Y.W. prepared the manuscript with the input of all authors. S.W. and M.T. discussed the results and contributed to the manuscript.

## Competing interests

The authors declare no competing interests.

**Fig. 1 | Magnetic configurations of skyrmion bags. a** Schematic representation of the FeGe microdevice. **b** Skyrmions with $Q$ = 1 and −1. **c** Representative examples of $S(n)$ and $S(n_1, S(n_2))$ skyrmion bags. The color coding indicates the magnetization orientation as defined in the colorbar: white represents an out-of-plane upward magnetization, while black represents an out-of-plane downward magnetization.

**Fig. 2 | Current-induced creation of skyrmion bags at zero magnetic fields.** Initial helix domains after applying a single high pulsed current. Blue and red boxes represent the closure helix domains topologically equivalent to skyrmions with $Q$ = 1 and −1, respectively. **b** Magnetic configuration after applying 10 pulsed currents on initial helix domains. **c**. Snapshots of creating skyrmion bags during the application of a series of pulsed currents. **d** Schematic formation of $S(n)$ bags because of the split of helix during the rush of moving individual skyrmions. **e** Schematic formation of $S(S(n))$ bags because of the split of helix during the rush of internal individual skyrmions. The arrows in **d** and **e** mark the moving orientations of individual skyrmions.

**Fig. 3 | Representative skyrmion bags captured during the cycle of current-induced helix-to-skyrmion transformations. a, b** Experimental observations (a) and corresponding simulations (b) of traditional $S(n)$ skyrmion bags, exhibiting a topological charge of $Q = n - 1$. **c, d** Experimental observations (c) and corresponding simulations (d) of nested $S(n_1, S(n_2))$ skyrmion bags with $Q = n_1 - n_2$. The experimental magnetizations are retrieved through a typical transport of intensity equation (TIE) analysis, while the simulated magnetizations are extracted from the middle layers of the skyrmion bags. The colors in (a) and (c) represent experimental in-plane magnetization according to the colorwheel in (a). The colors in (b) and (d) represent simulated magnetizations according to the colorwheel in (b).

**Fig. 4 | Probability for electrical creation of skyrmion bags. a** Counts of magnetic skyrmion bags, $N$, observed over 50 independent cycles, during the transition from a disordered helical state to a skyrmion lattice. The generation probability is calculated as $p = N/50 \times 100\%$, representing the likelihood of skyrmion bag formation per current cycle. Each current cycle consists of a single high-current-density pulse followed by 150 low-current-density pulses. The statistical data were obtained from three devices with similar dimensions: length ~4 μm, width ~3 μm, and thickness ~100 nm. Error bars represent the standard deviation across 3 devices. **b** Simulated total energy difference $\Delta E$ between skyrmion bags and the $S(1)$ bag embedded in the skyrmion lattice.

**Fig. 5 | Current-induced collapse process of interior skyrmion tubes of a nested $S(0, S(2))$ skyrmion bag.** Experimental Fresnel contrasts, simulated Fresnel contrasts, and simulated 3D magnetization isosurfaces during the collapse process. **a-c** An initial nested $S(0, S(2))$ skyrmion bag. **d-f** A nested skyrmion bag containing two merging skyrmion tubes was obtained after applying 6 current pulses. **g-i** A nested $S(0, S(1))$ skyrmion bag containing one tube obtained after applying 7 current pulses. **j-l** a nested skyrmion bag containing 1 bobber obtained after applying 13 current pulses. **m-o** $S(1)$ skyrmion bag obtained after applying 13 current pulses. The iso-surface corresponds to zero out-of-plane magnetization. The color scheme in the isosurface illustrates the in-plane magnetization, based on the colorwheel. Defocused distance of Fresnel contrasts, −500 μm. **p, q** Experimental (p) and simulated (q) Fresnel contrasts along the corresponding white dashed lines.

## Additional information

Supplementary information is available for this paper.

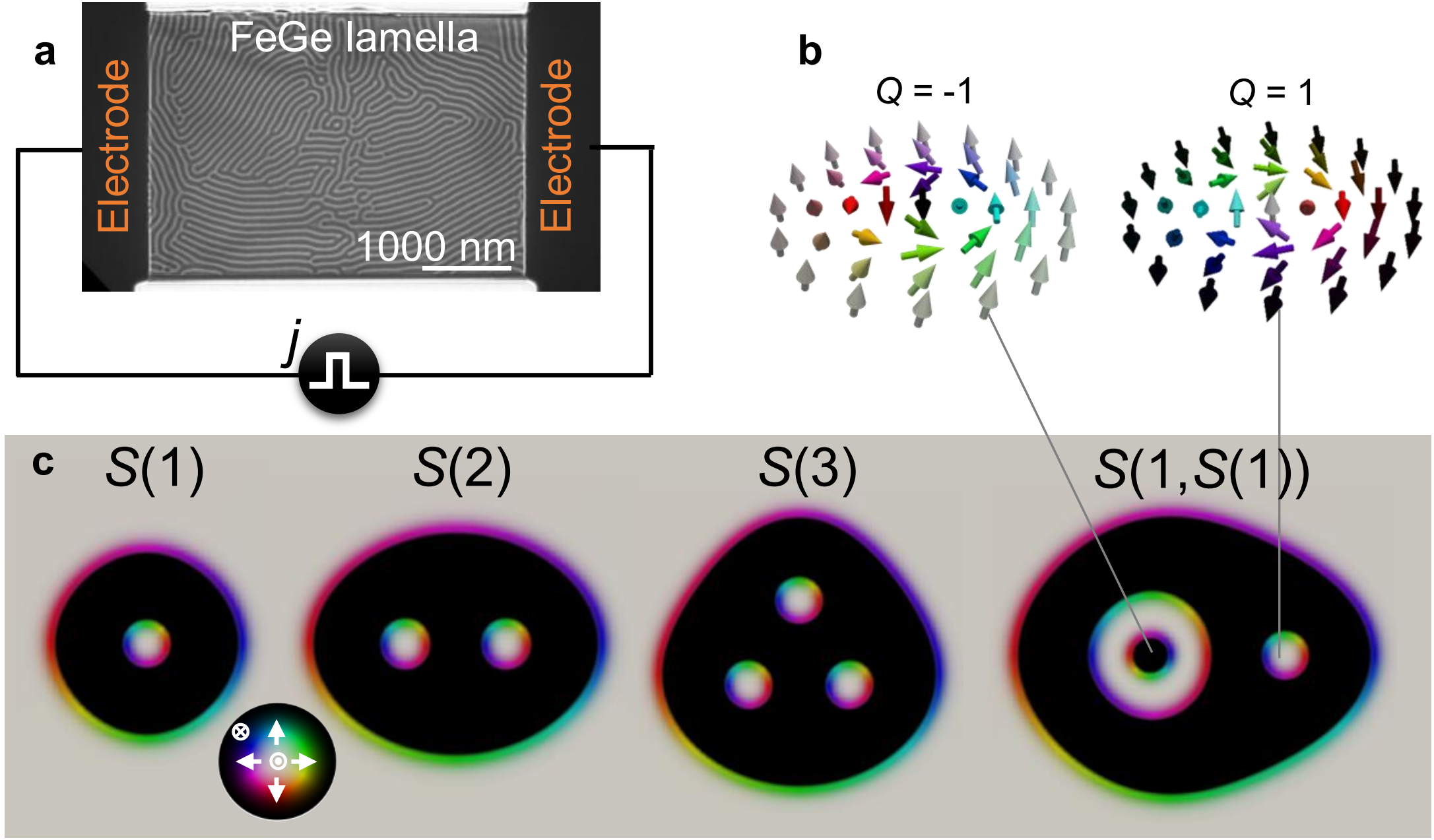
a
FeGe lamella
Electrode
Electrode
1000 nm
j
b
Q = -1
Q = 1
c
S(1)
S(2)
S(3)
S(1,S(1))

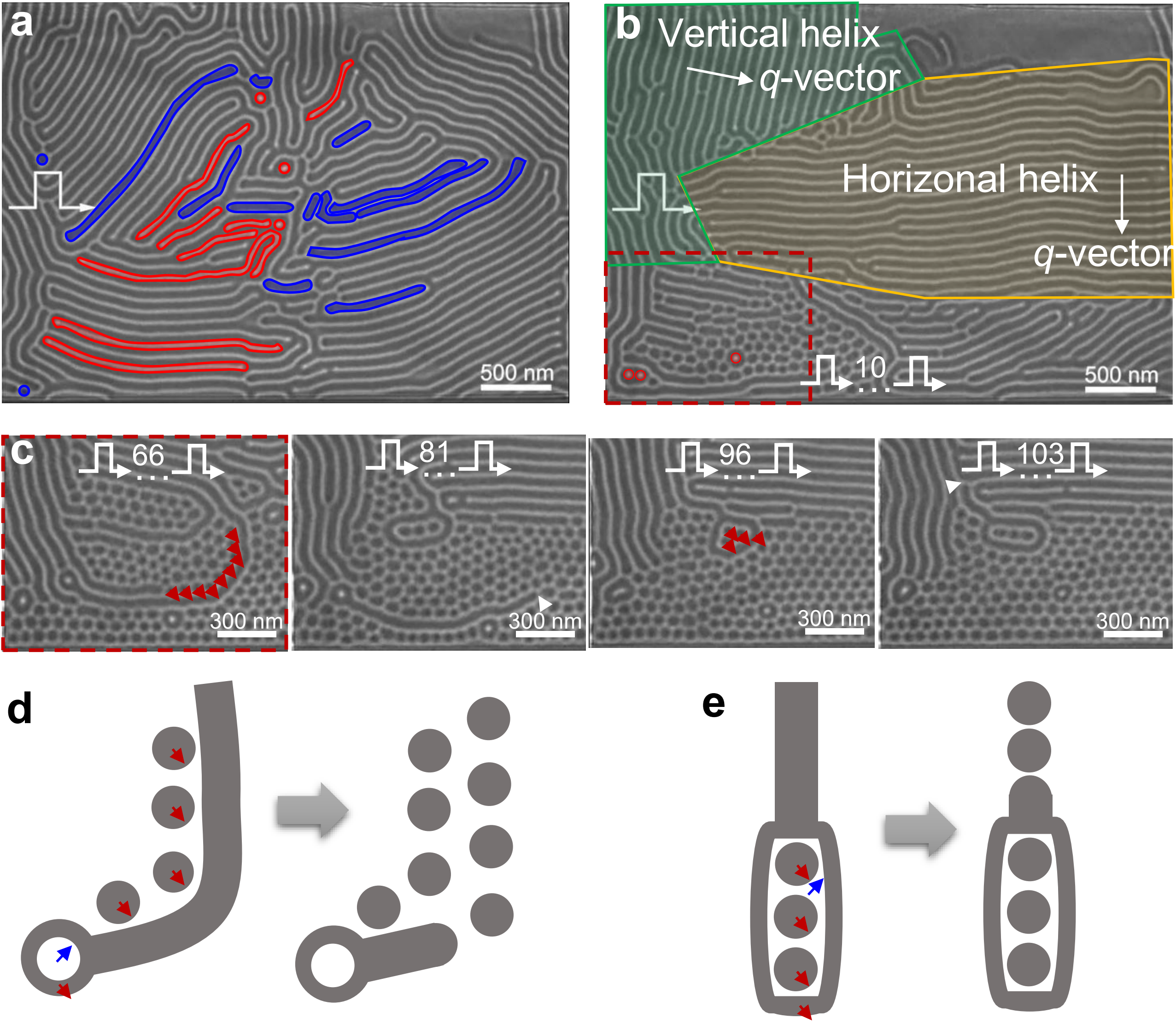
a
500 nm
b
Vertical helix
q-vector
Horizonal helix
q-vector
10
500 nm
c
66
300 nm
81
300 nm
96
300 nm
103
300 nm
d
e

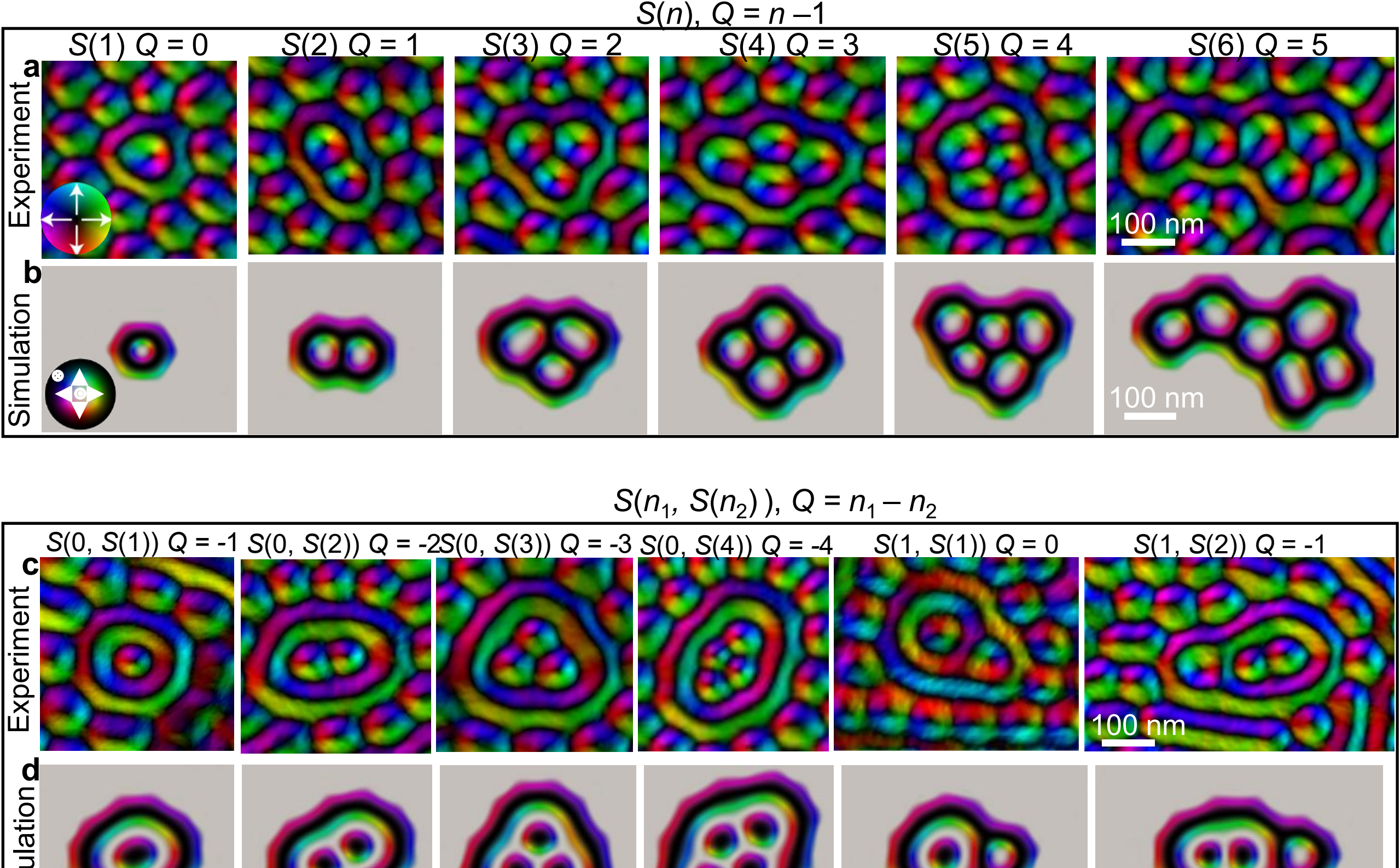

$S(n)$, $Q = n - 1$
$S(1)$ $Q = 0$
$S(2)$ $Q = 1$
$S(3)$ $Q = 2$
$S(4)$ $Q = 3$
$S(5)$ $Q = 4$
$S(6)$ $Q = 5$
a
Experiment
b
Simulation
100 nm
100 nm
$S(n_1, S(n_2))$, $Q = n_1 - n_2$
$S(0, S(1))$ $Q = -1$
$S(0, S(2))$ $Q = -2$
$S(0, S(3))$ $Q = -3$
$S(0, S(4))$ $Q = -4$
$S(1, S(1))$ $Q = 0$
$S(1, S(2))$ $Q = -1$
c
Experiment
d
Simulation
100 nm
100 nm

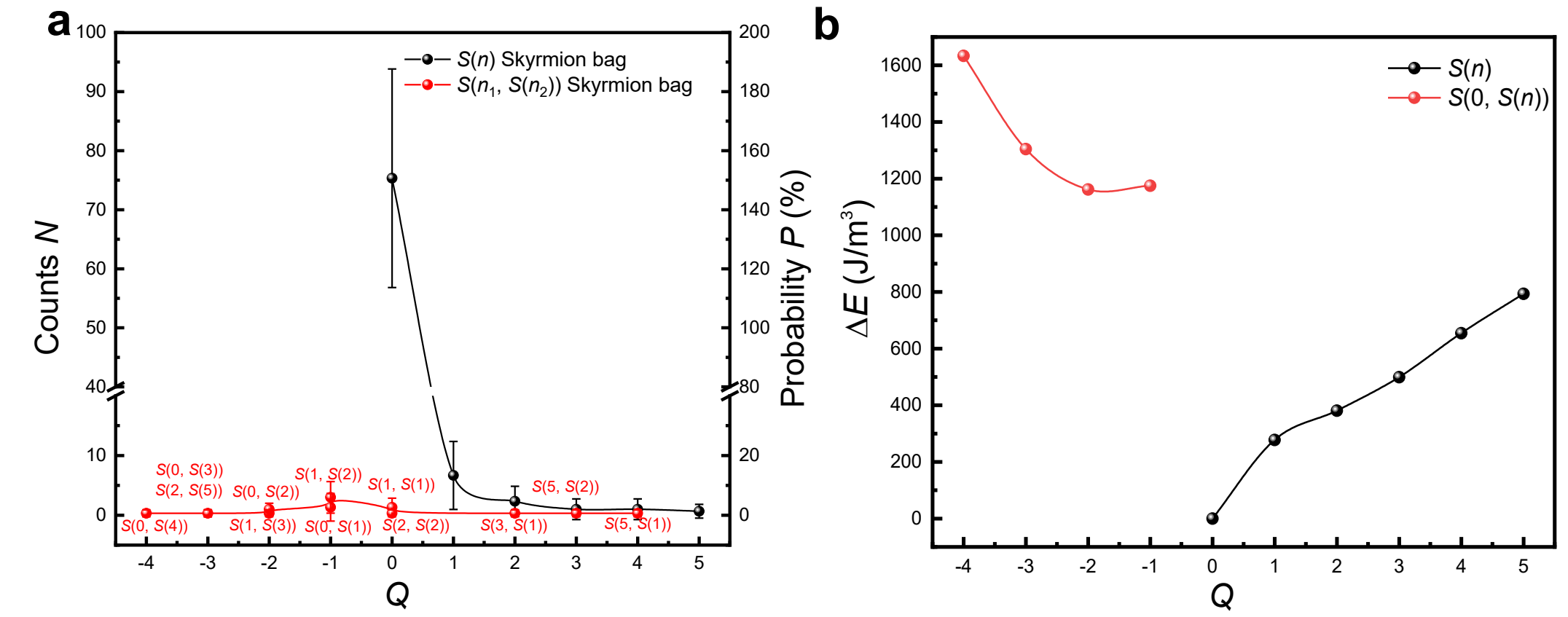

a
b
Counts $N$
Probability $P$ (%)
$Q$
$S(n)$ Skyrmion bag
$S(n_1, S(n_2))$ Skyrmion bag
$S(0, S(3))$
$S(2, S(5))$
$S(0, S(2))$
$S(1, S(2))$
$S(1, S(1))$
$S(5, S(2))$
$S(0, S(4))$
$S(1, S(3))$
$S(0, S(1))$
$S(2, S(2))$
$S(3, S(1))$
$S(5, S(1))$
$\Delta E$ (J/m$^3$)
$S(n)$
$S(0, S(n))$

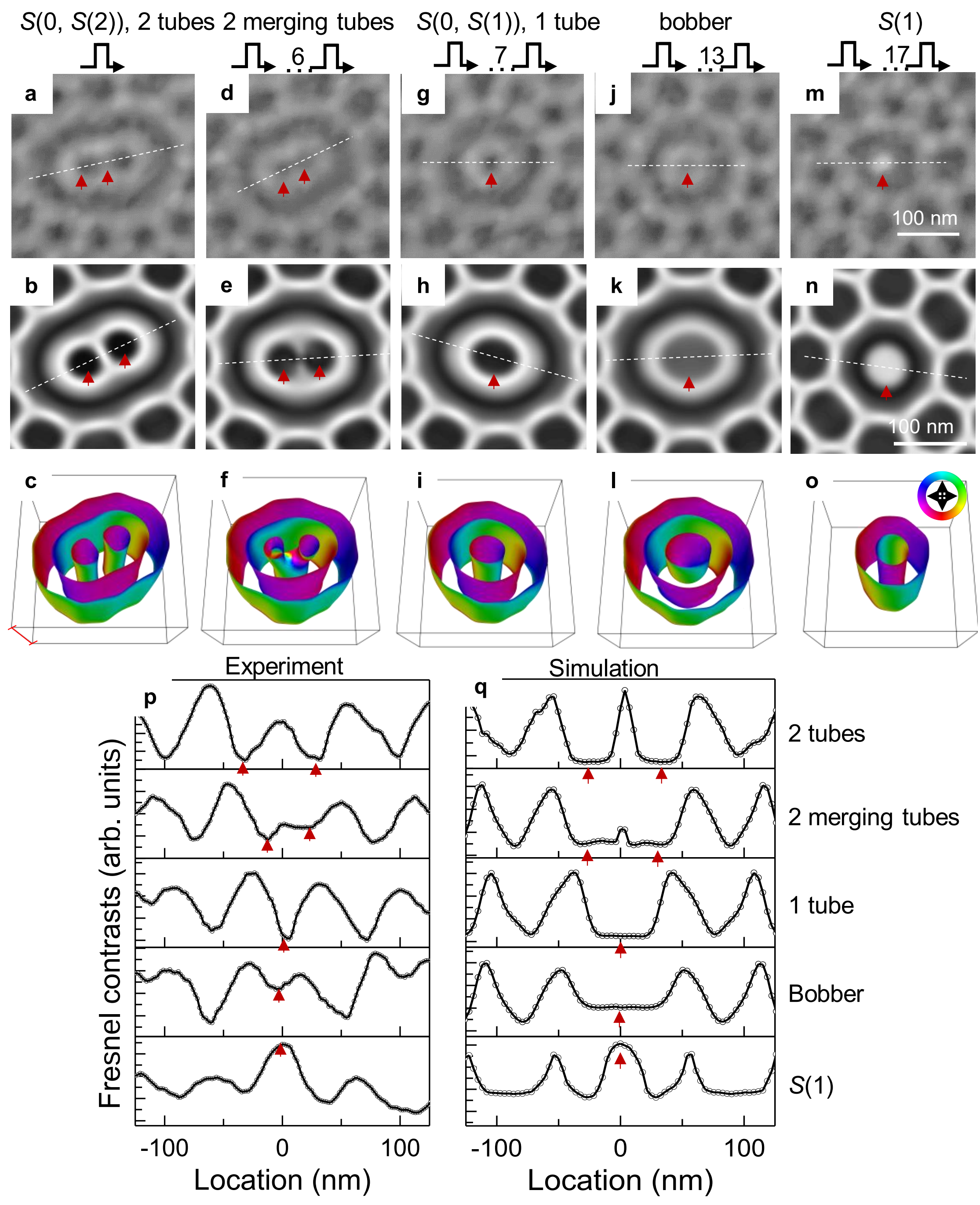

S(0, S(2)), 2 tubes
2 merging tubes
S(0, S(1)), 1 tube
bobber
S(1)
6
7
13
17
a
b
c
d
e
f
g
h
i
j
k
l
m
n
o
100 nm
100 nm
Experiment
Simulation
p
q
Fresnel contrasts (arb. units)
2 tubes
2 merging tubes
1 tube
Bobber
S(1)
-100
0
100
Location (nm)
-100
0
100
Location (nm)

# Supporting Information

## Current-induced creation and dynamics of embedded magnetic skyrmion bags

Yaodong Wu[1], Jialiang Jiang[2, 3*], Meng Shi[3], Shouguo Wang[4], Mingliang Tian[1,2], Haifeng Du[3*], and Jin Tang[2, 3*]

[1]School of Physics and Materials Engineering, Hefei Normal University, Hefei 230601, China

[2]State Key Laboratory of Opto-Electronic Information Acquisition and Protection Technology, School of Physics, Anhui University, Hefei, 230601, China

[3]Anhui Provincial Key Laboratory of Low-Energy Quantum Materials and Devices, High Magnetic Field Laboratory, HFIPS, Chinese Academy of Sciences, Hefei 230031, China

[4]Anhui Key Laboratory of Magnetic Functional Materials and Devices, School of Materials Science and Engineering, Anhui University, Hefei 230601, China

*Corresponding author: jintang@ahu.edu.cn; duhf@hmfl.ac.cn; jjl2024@ahu.edu.cn

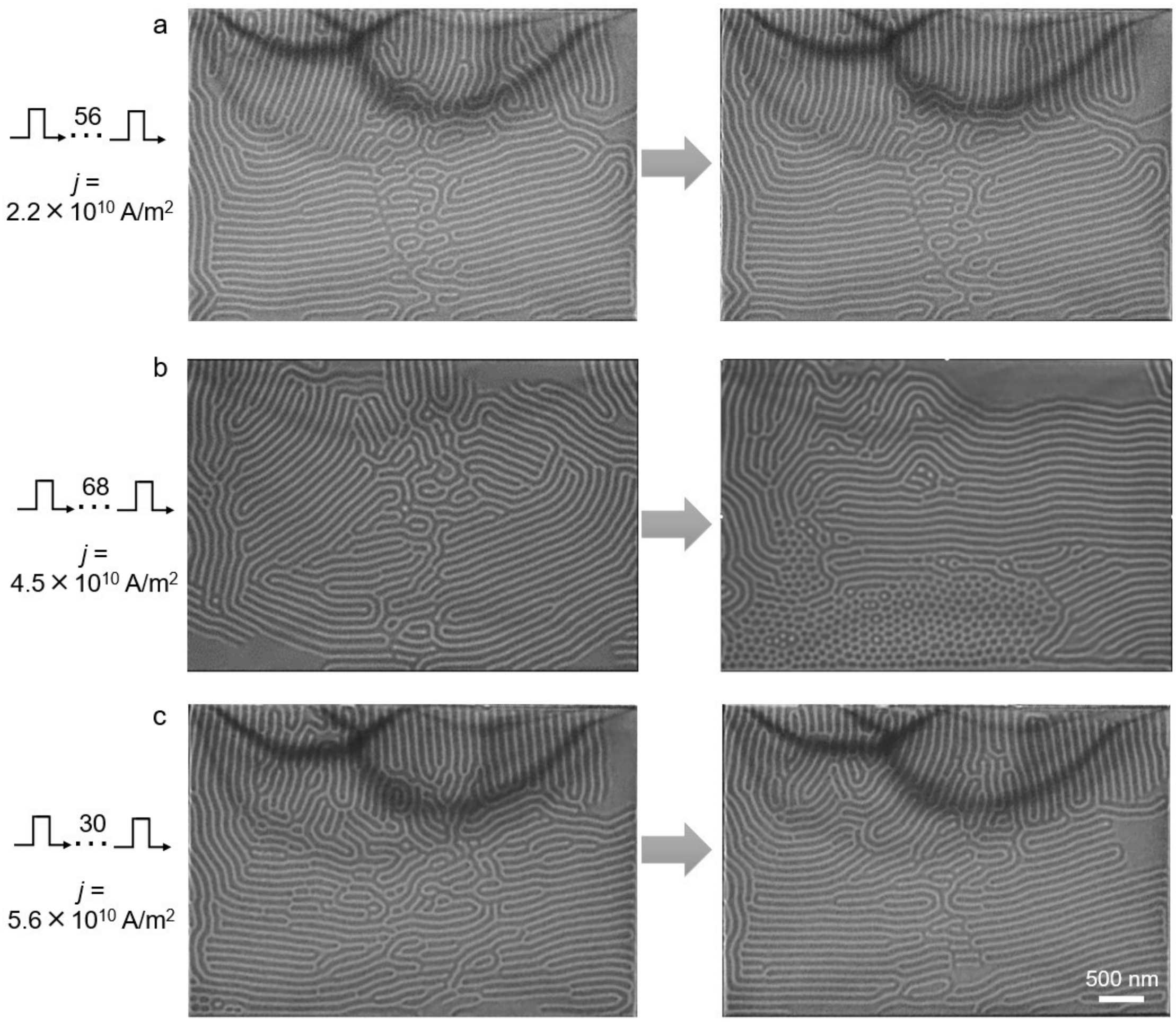

**Supplementary Fig. 1 | Current-driven dynamics from the initial helix.** **a** $j = 2.2 \times 10^{10}$ A/m$^2$. **b** $j = 4.5 \times 10^{10}$ A/m$^2$. **c** $j = 5.9 \times 10^{10}$ A/m$^2$. Pulse duration, 70 ns.

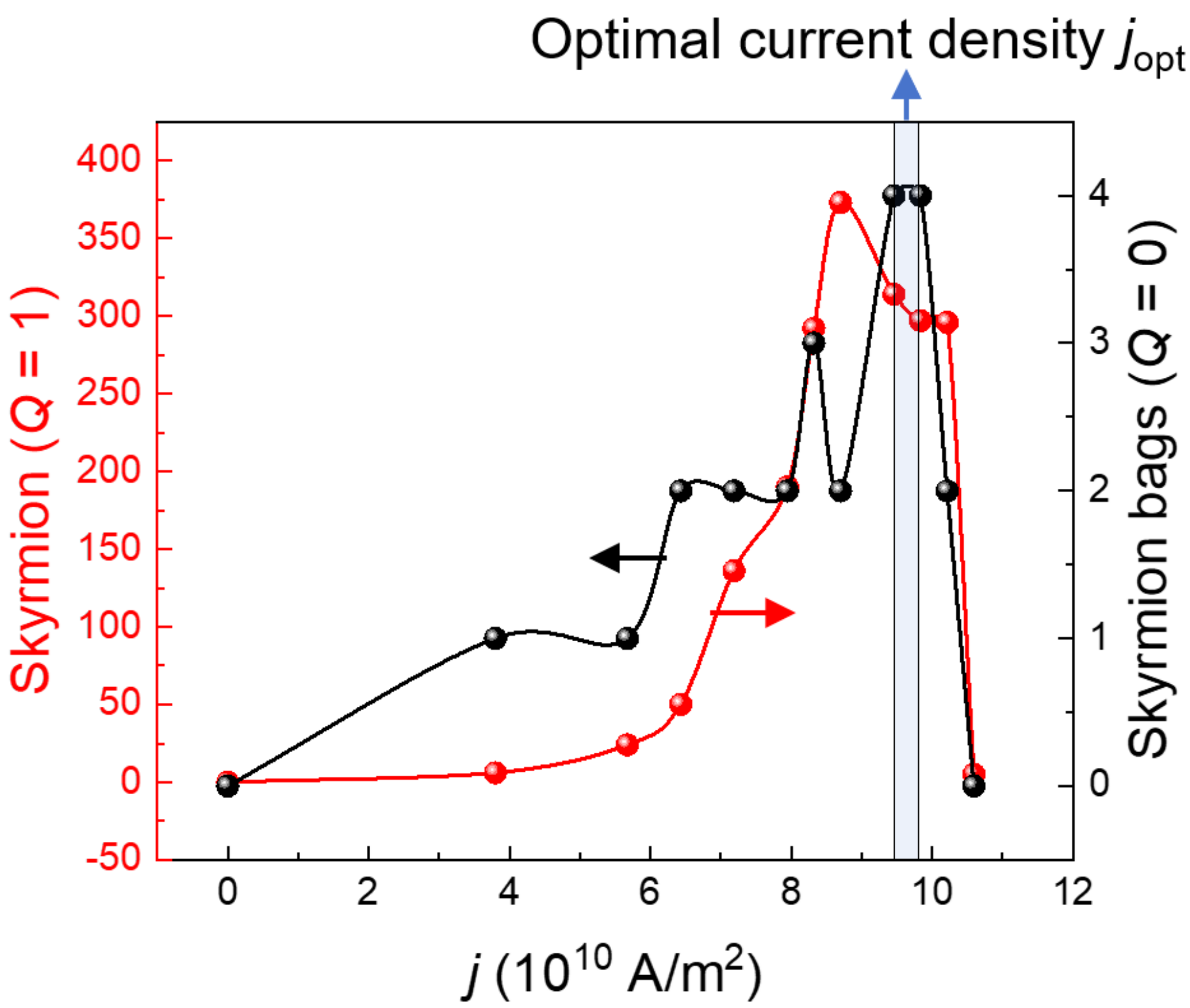

**Supplementary Fig. 2 | The maximum skyrmion and $Q$ = 0 skyrmion bag counts as a function of current density induced by current during each current cycle.** Temperature, 100 K. Pulse duration, 20 ns.

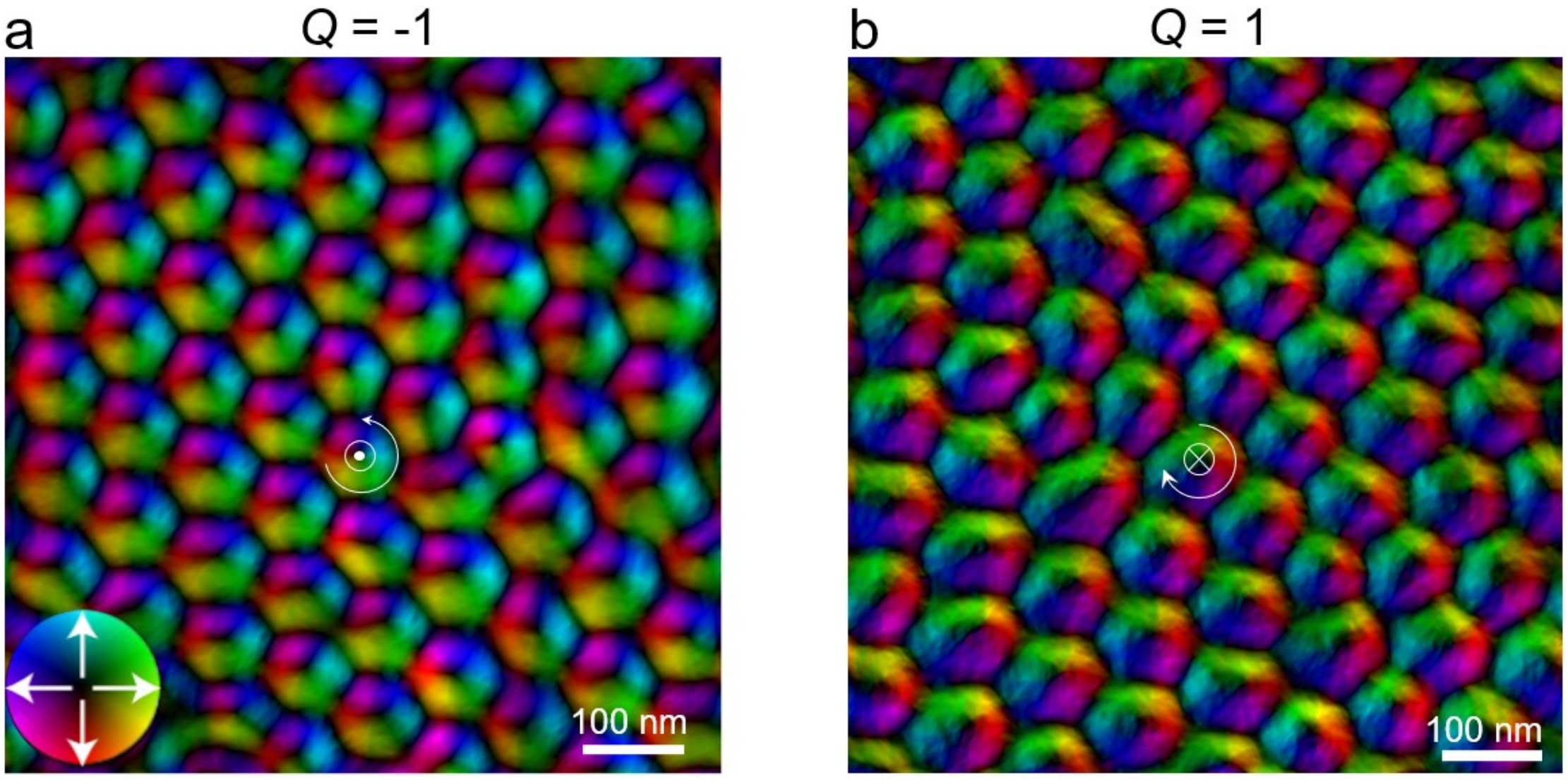

**Supplementary Fig. 3 | Zero-field magnetic skyrmion lattices. a, b** In-plane magnetization mapping of skyrmions with $Q = 1$ (a) and $-1$ (b) retrieved from TIE.

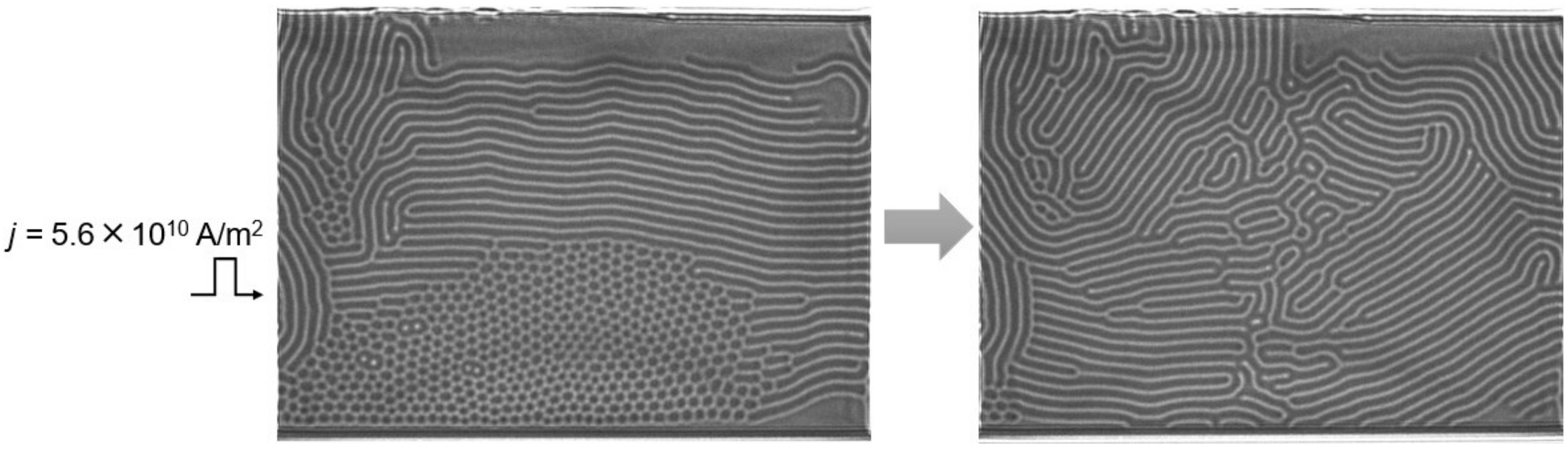

**Supplementary Fig. 4 | Initial helix obtained by applying a single high current with a density of $5.6 \times 10^{10}$ A/m$^2$.**

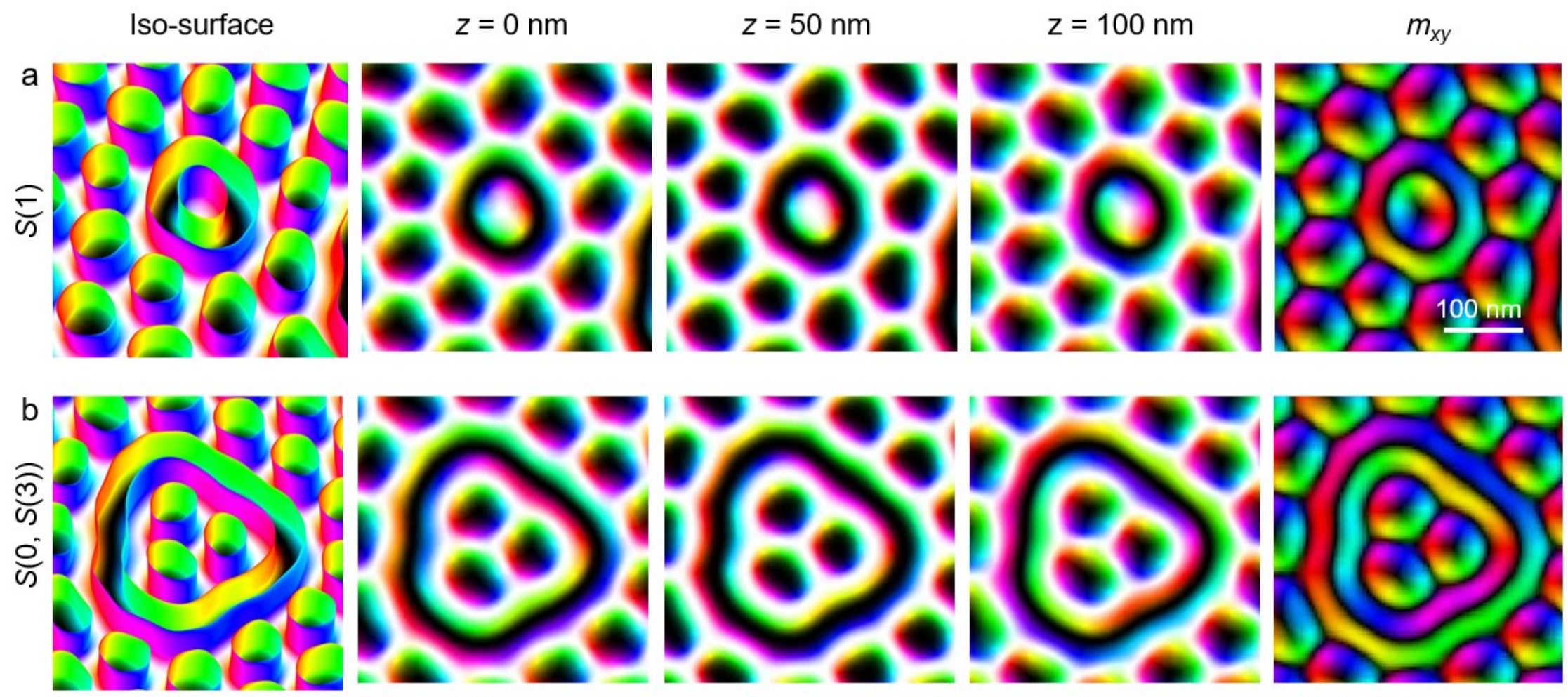

**Supplementary Fig. 5 | 3D configuration of skyrmion bags. a, b** 3D iso-surfaces for $m_z = 0$, magnetization at the top layer $z = 0$ nm, magnetization at the middle layer $z = 50$ nm, magnetization at the bottom layer, and in-plane average magnetization for the $S(1)$ (a) and $S(0, S(3))$ (b) bags.

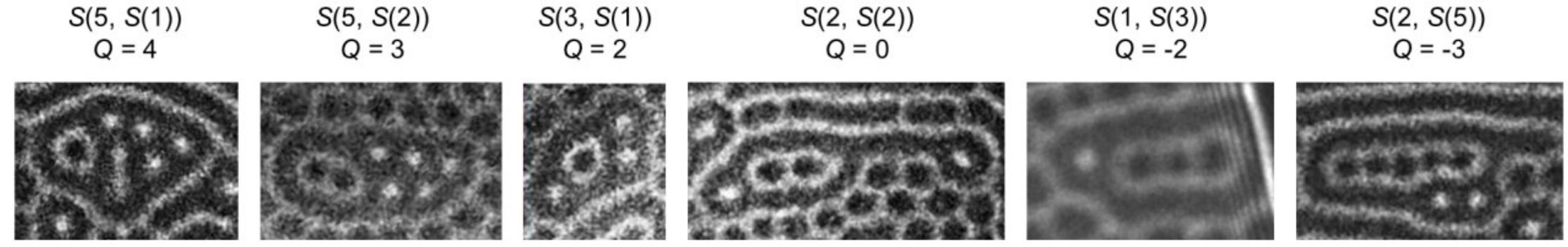

**Supplementary Fig. 6 | Experimental observations of nested magnetic skyrmion bags at a zero magnetic field.** De-focused Fresnel contrasts of *S*(5, *S*(1)), *S*(1, *S*(1)), *S*(5, *S*(2)), *S*(3, *S*(1)), *S*(2, *S*(2)), *S*(1, *S*(3)), and *S*(2, *S*(5)) nested skyrmion bags. Defocused distance, −500 μm.

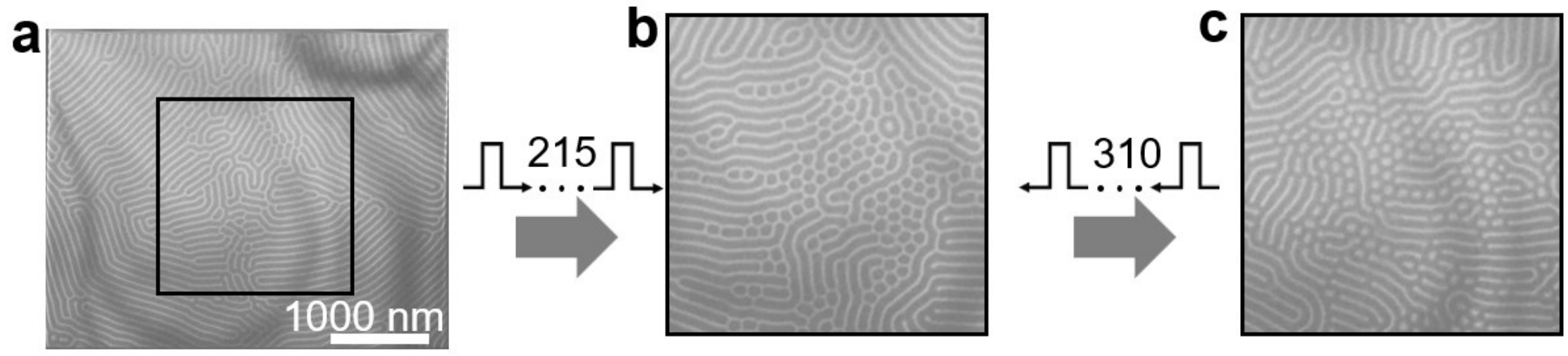

**Supplementary Fig. 7 | Skyrmion polarity reversal by reversing the current orientation.** **a** Initial domain states. **b** Current-induced skyrmions and skyrmion bags after applying 215 current pulses with a density of $+8.4 \times 10^{10}$ A m$^{-2}$. **c** Subsequent state after applying 310 current pulses with a density of $-8.4 \times 10^{10}$ A m$^{-2}$. The Fresnel images in (b) and (c) correspond to the region indicated by the black box in (a). Defocus distance: −500 μm.

**Supplementary Table 1 | Parameters for Multiphysics field simulations**.

| Name | Parameter | Value | Unit | Notes |
|---|---|---|---|---|
| **FeGe** | Density ($\rho_m$) | 8220 | kg $m^{-3}$ | Ref. [1] |
| | Heat capacity ($C_p$) | 363 | J $kg^{-1}$ $K^{-1}$ | Ref. [2] |
| | Thermal conductivity ($k$) | 3.39 | W $m^{-1}$ $K^{-1}$ | Ref. [2] |
| | Resistivity ($\rho$) | $(1+0.00948(T\text{-}100)) \times 80.87$ | μΩ cm | Ref. [2] |
| | Heat transfer coefficient ($h$) | $5 \times 10^6$ | W $m^{-2}$ $K^{-1}$ | Ref. [2] |
| **I-Pt** | Density ($\rho_m$) | 21450 | kg $m^{-3}$ | Ref. [3] |
| | Heat capacity ($C_p$) | 132 | J $kg^{-1}$ $K^{-1}$ | Ref. [3] |
| | Thermal conductivity ($k$) | 71.6 | W $m^{-1}$ $K^{-1}$ | Ref. [3] |
| | Resistivity ($\rho$) | $(1+0.00385(T\text{-}100)) \times 2.71$ | μΩ cm | Ref. [4] |
| | Heat transfer coefficient ($h$) | $5 \times 10^6$ | W $m^{-2}$ $K^{-1}$ | Ref. [2] |

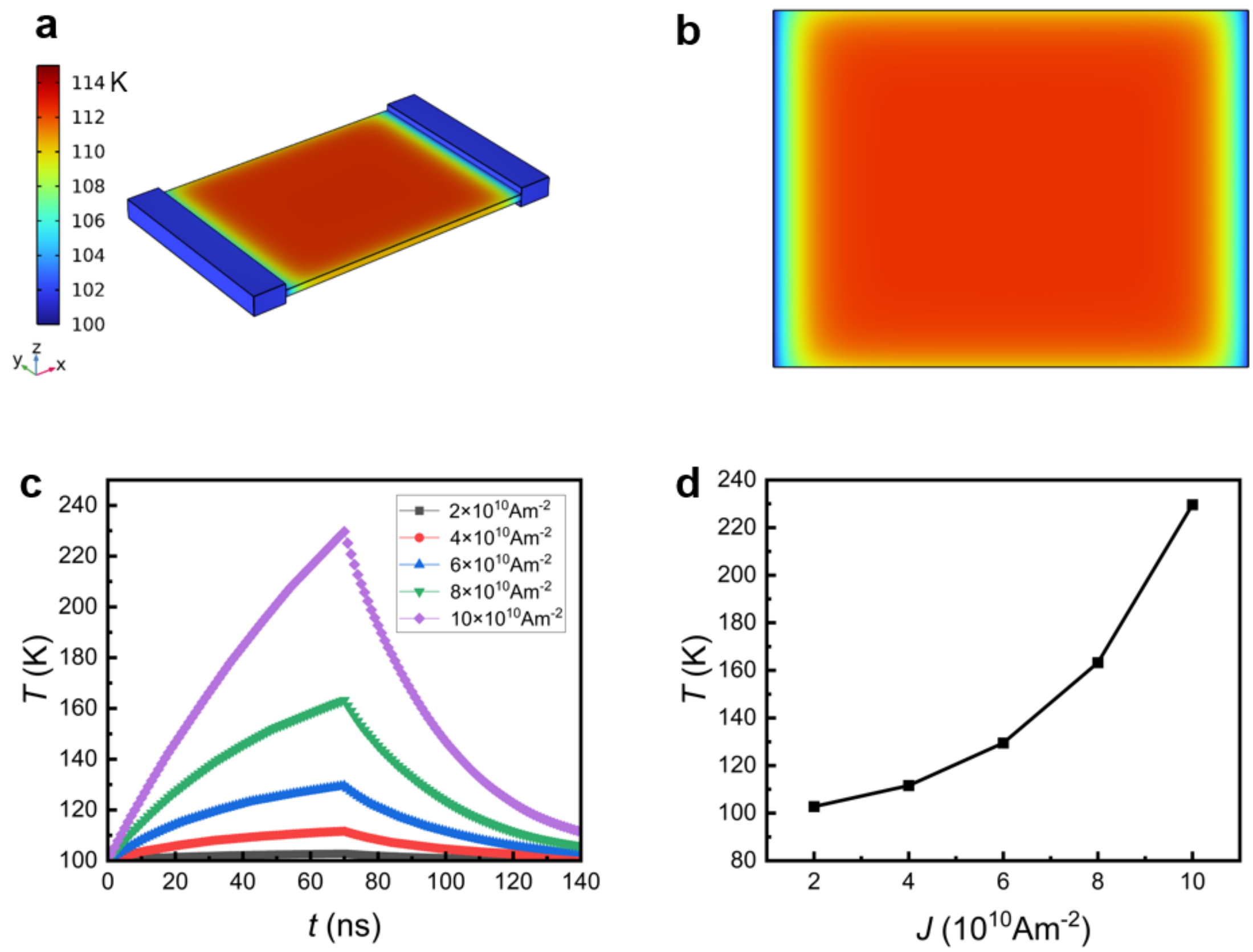

**Supplementary Fig. 8 | Multiphysics field simulation of current-induced Joule thermal heating.** **a** Simulated temperature distribution after a 70-ns current pulse with a current density of $4\times10^{10}$ A m$^{-2}$. **b** Enlarged view of the temperature distribution in the device area. **c** Temperature at the center of the device as a function of time during the pulse for different current densities. **d** Maximum temperature rise versus current density for 70-ns pulses.

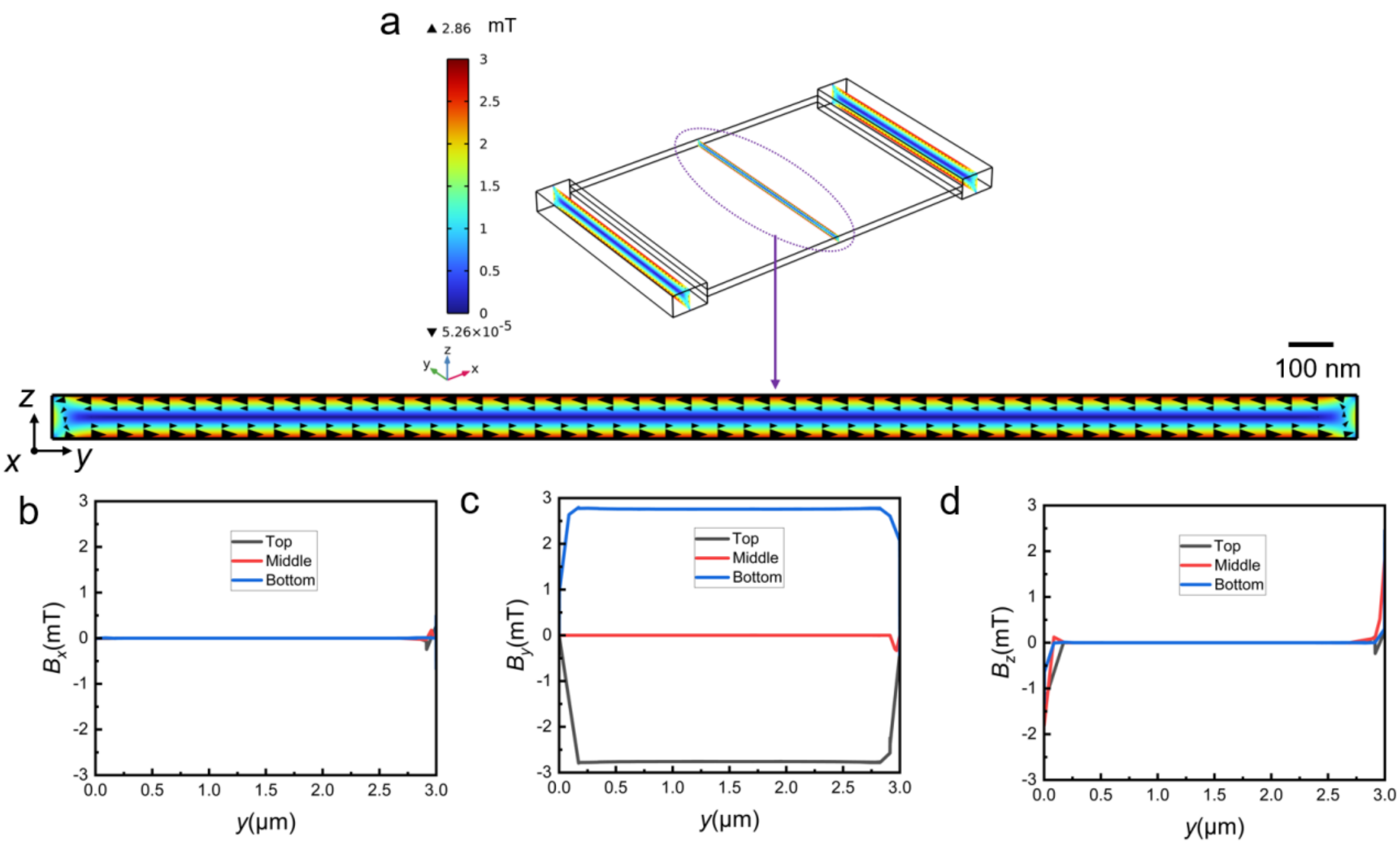

**Supplementary Fig. 9 | Simulated Oersted field distribution of a 100-nm-thick FeGe microdevice at a current density of 4.5×10$^{10}$ A·m$^{-2}$ applied along the *x* axis. a** The Oersted field distribution in the middle layer. The color represents the magnitude of the magnetic flux density. **b-d** Dependence of the Oersted field components on position *y* at the middle layer: $B_x$ (b), $B_y$ (c), and $B_z$ (d).

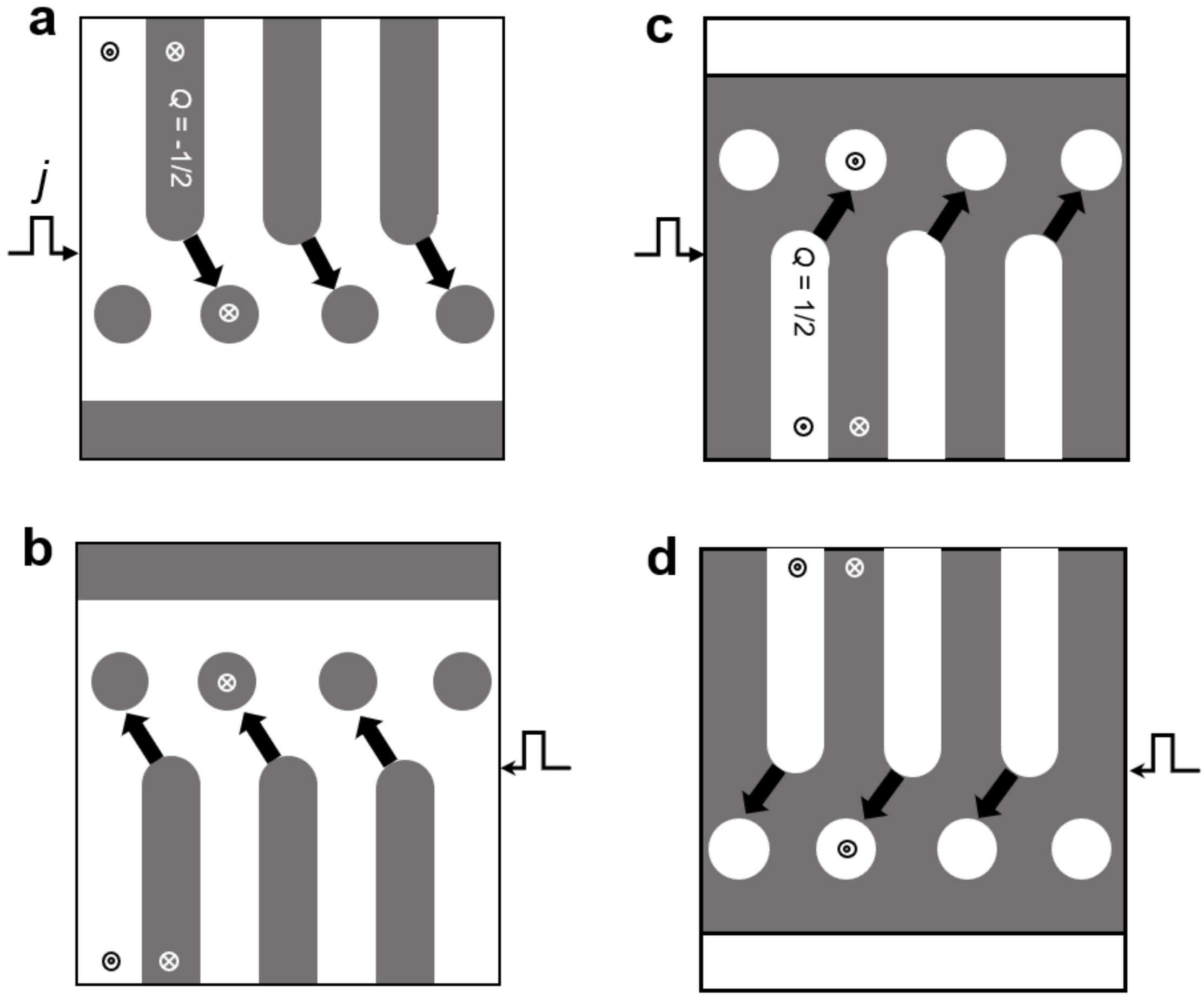

**Supplementary Fig. 10 | Schematic skyrmion generation rule based on force and symmetry analysis.** Four cases are considered depending on the helix pinning ends and current orientation. Black and white represent out-of-plane magnetization down and up, respectively. Black arrows indicate the motion direction of helix ends, equivalent to merons driven by STT. **a** vertical helix pinned at the top ends, positive current; **b** pinned at the bottom ends, positive current; **c** pinned at the bottom ends, negative current; **d** pinned at the top ends, negative current. Starting from (a), other cases in (b), (c), and (d) can be obtained by applying time-reversal and/or combined rotation of spin and space by 180 degrees to (a).

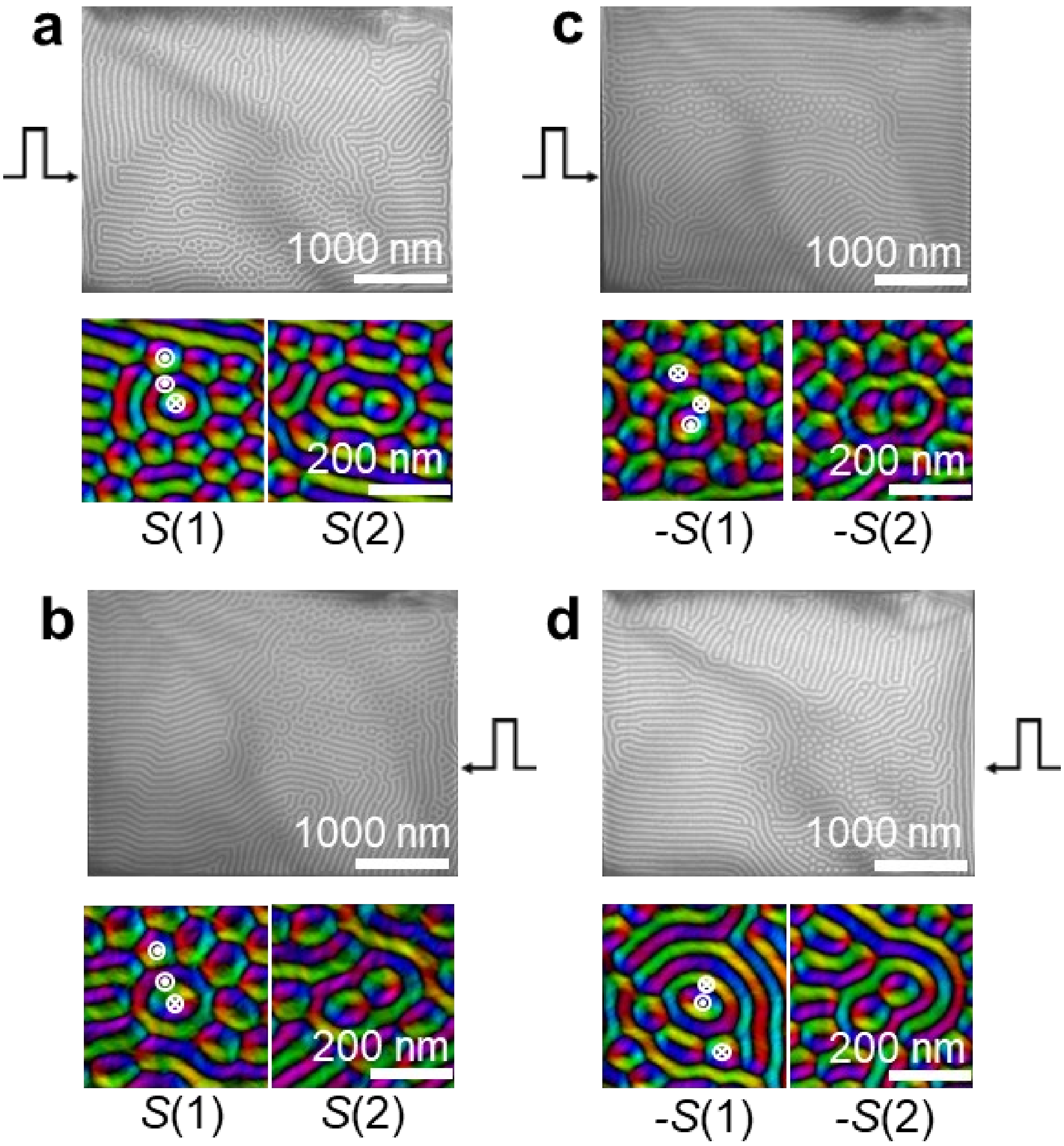

**Supplementary Fig. 11 | Experimental demonstration of the skyrmion generation rule based on force and symmetry analysis.** Fresnel images were taken at a defocus distance of −500 μm, revealing the pinning ends of the vertical helix. Representative in-plane magnetic configurations of skyrmion bags and skyrmions are shown for four cases, depending on the pinning location and current polarity: **a** vertical helix pinned at the top ends, positive current; **b** pinned at the bottom ends, positive current; **c** pinned at the bottom ends, negative current; **d** pinned at the top ends, negative current.

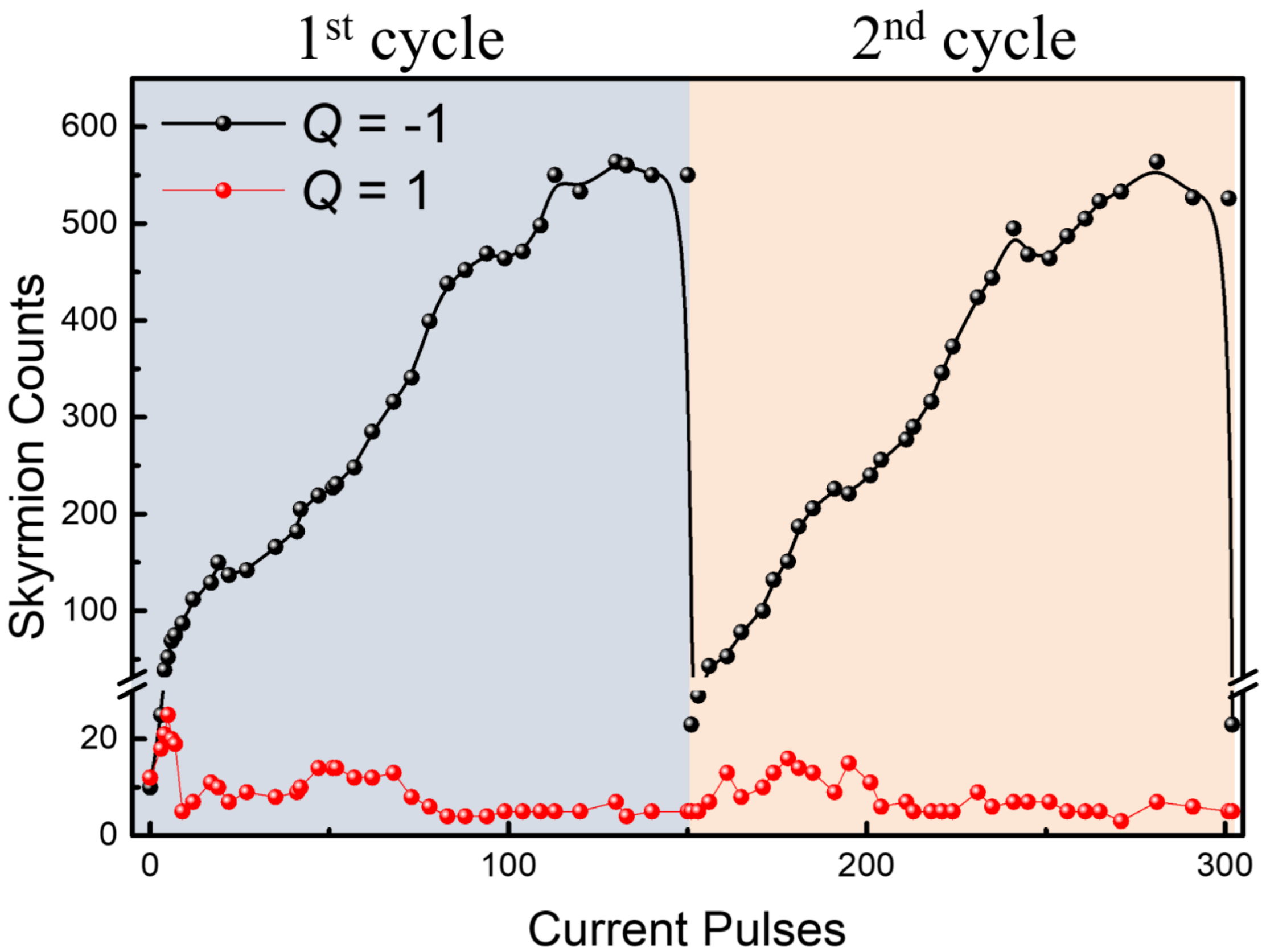

**Supplementary Fig. 12 | Count of $Q$ = 1 and −1 skyrmions during each current cycle.** Each current cycle consists of a single high-current pulse with a density of $5.5 \times 10^{10}$ A/m², followed by 150 low-current pulses with a density of $4.5 \times 10^{10}$ A/m². The pulse duration is set to 70 ns with a frequency of 1 Hz.

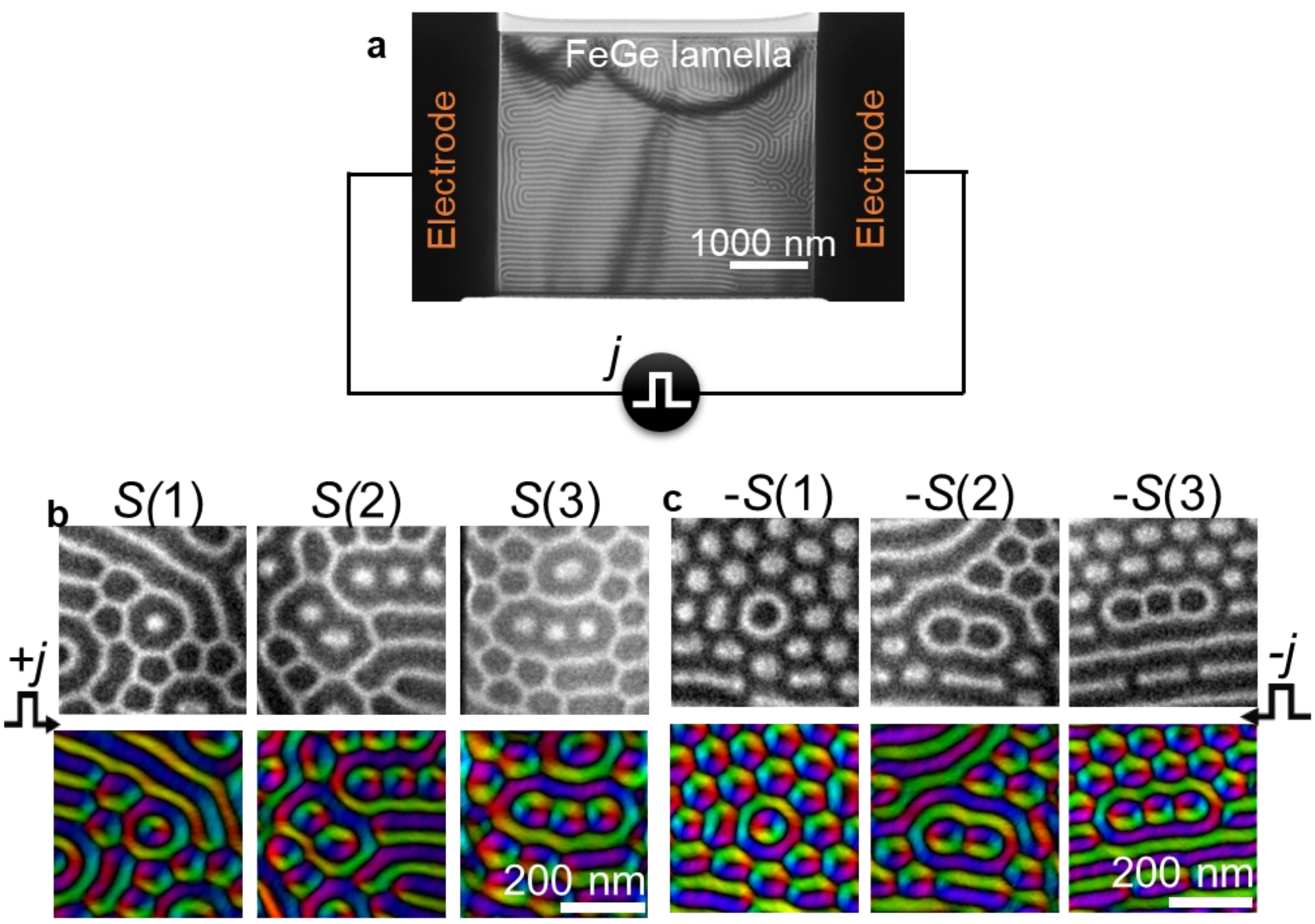

**Supplementary Fig. 13 | Experimental observation of skyrmion bag in an 80-nm thick FeGe device (#2) at 100 K and zero field.** (a) Overview of the device. (b) Representative skyrmion bags induced by positive currents. (c) Representative skyrmion bags induced by negative currents.

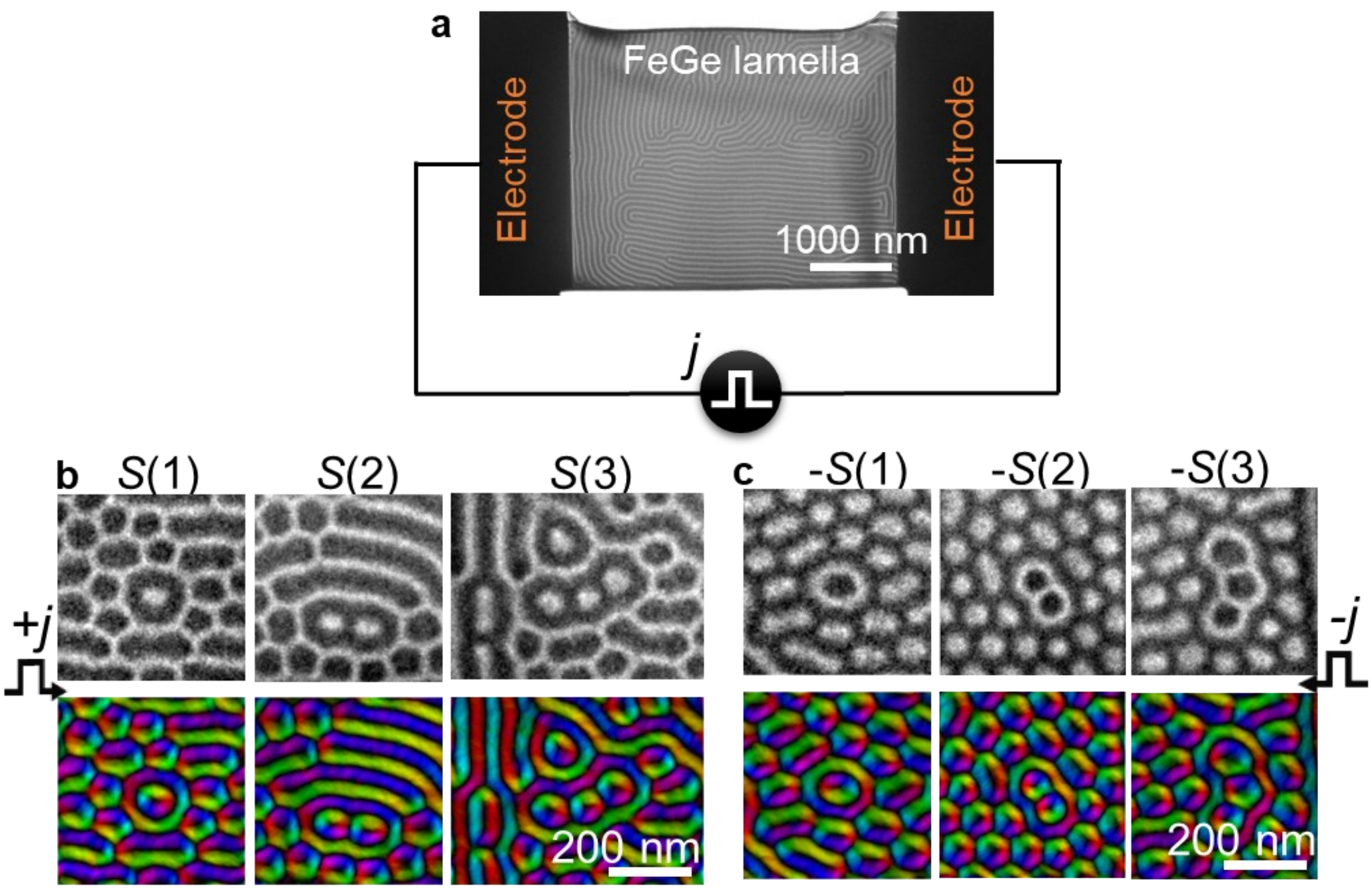

**Supplementary Fig. 14 | Experimental observation of skyrmion bag in a 120-nm thick FeGe device (#3) at 100 K and zero field.** (a) Overview of the device. (b) Representative skyrmion bags induced by positive currents. (c) Representative skyrmion bags induced by negative currents.

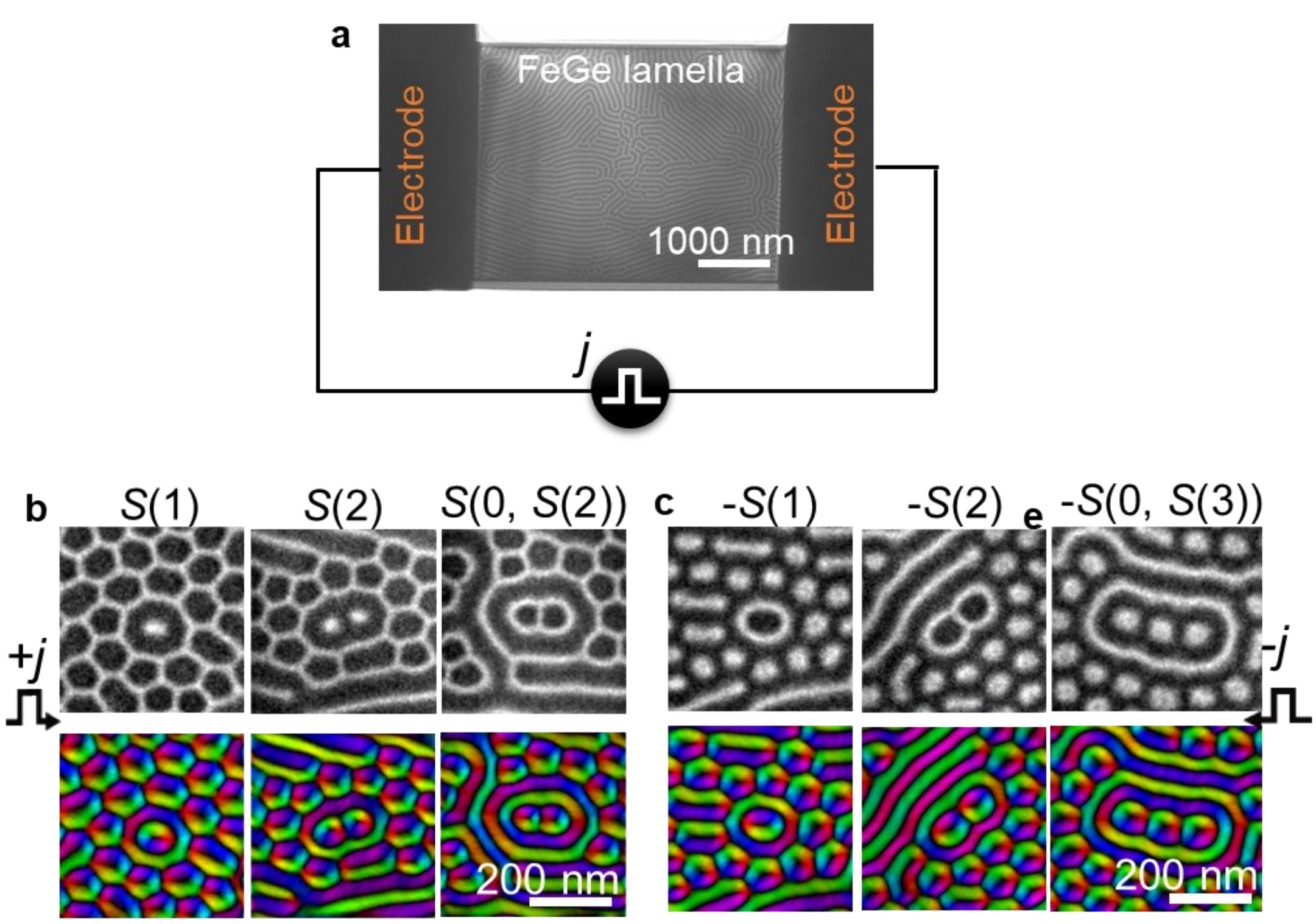

**Supplementary Fig. 15 | Experimental observation of skyrmion bag in a 130-nm thick FeGe device (#4) at 100 K and zero field.** (a) Overview of the device. (b) Representative skyrmion bags induced by positive currents. (c) Representative skyrmion bags induced by negative currents.

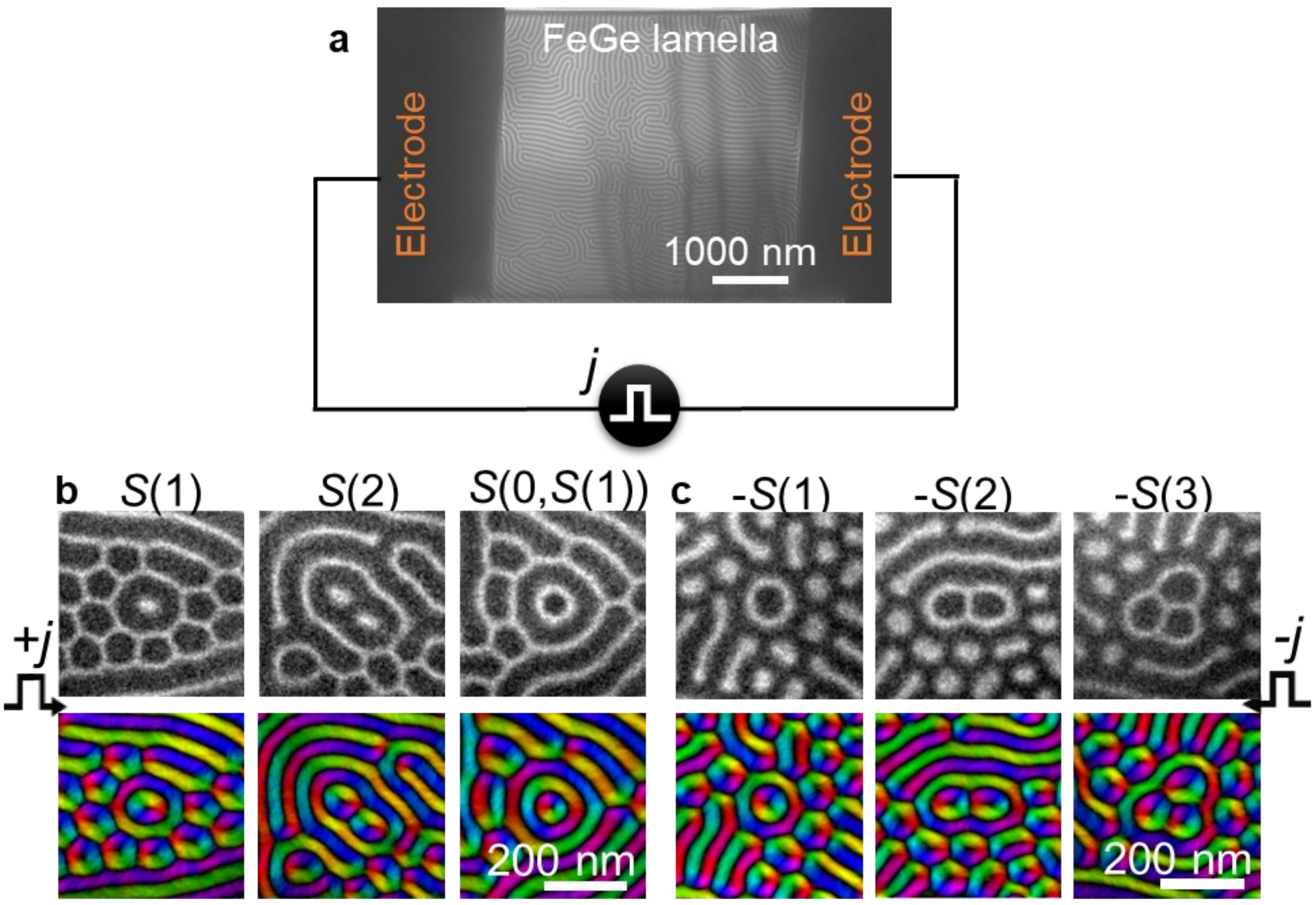

**Supplementary Fig. 16 | Experimental observation of skyrmion bag in a 150-nm thick FeGe device (#5) at 100 K and zero field.** (a) Overview of the device. (b) Representative skyrmion bags induced by positive currents. (c) Representative skyrmion bags induced by negative currents.

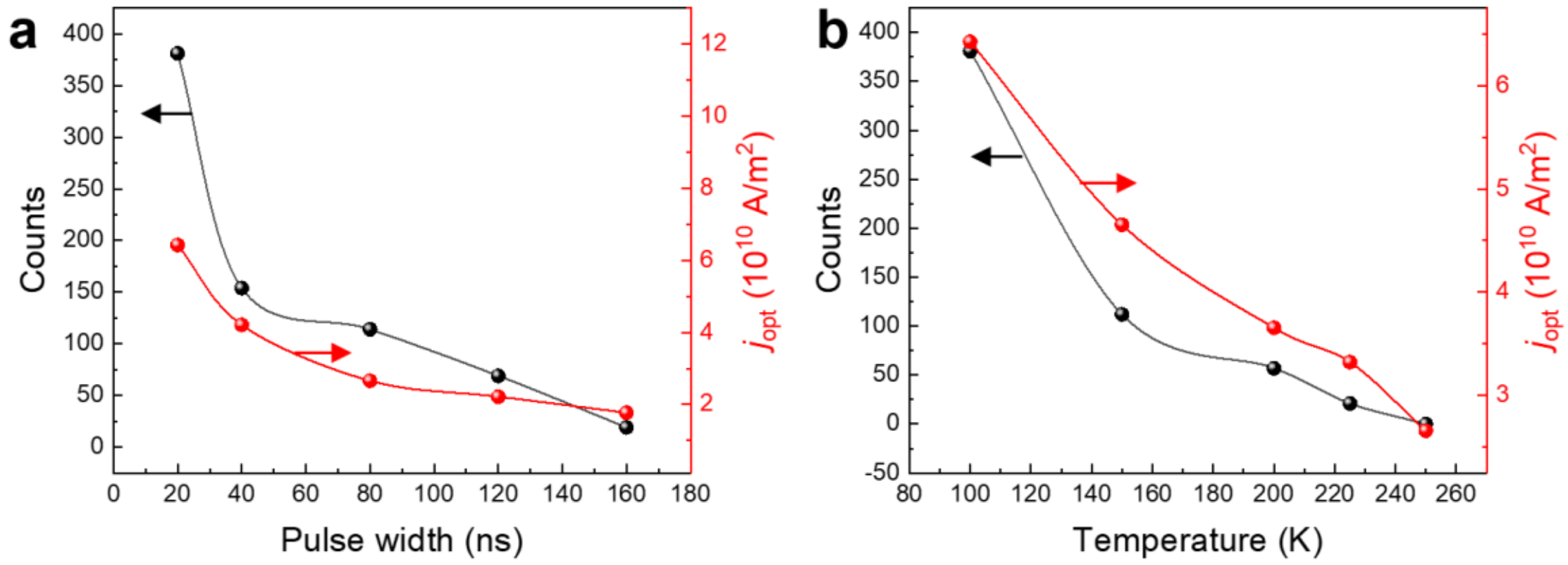

**Supplementary Fig. 17 | Effect of pulse width and temperature on the current-induced creation of skyrmion bags. a** The $S(1)$ skyrmion bag emerging in 50 current cycles as a function of pulse width. The current during the cycle is chosen as the optimal current density. **b** The $S(1)$ skyrmion bag emerging in 50 current cycles as a function of temperature. The current during the cycle is chosen as the optimal current density. The pulse width is 20 ns.

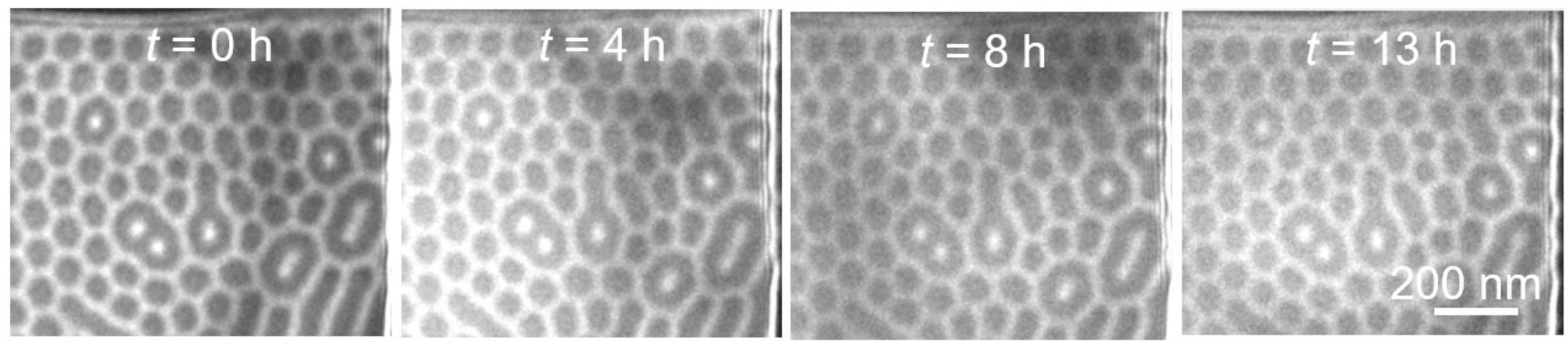

**Supplementary Fig. 18 | Skyrmion bag stability over time under constant conditions** (zero applied current, 95 K, and zero external magnetic field).

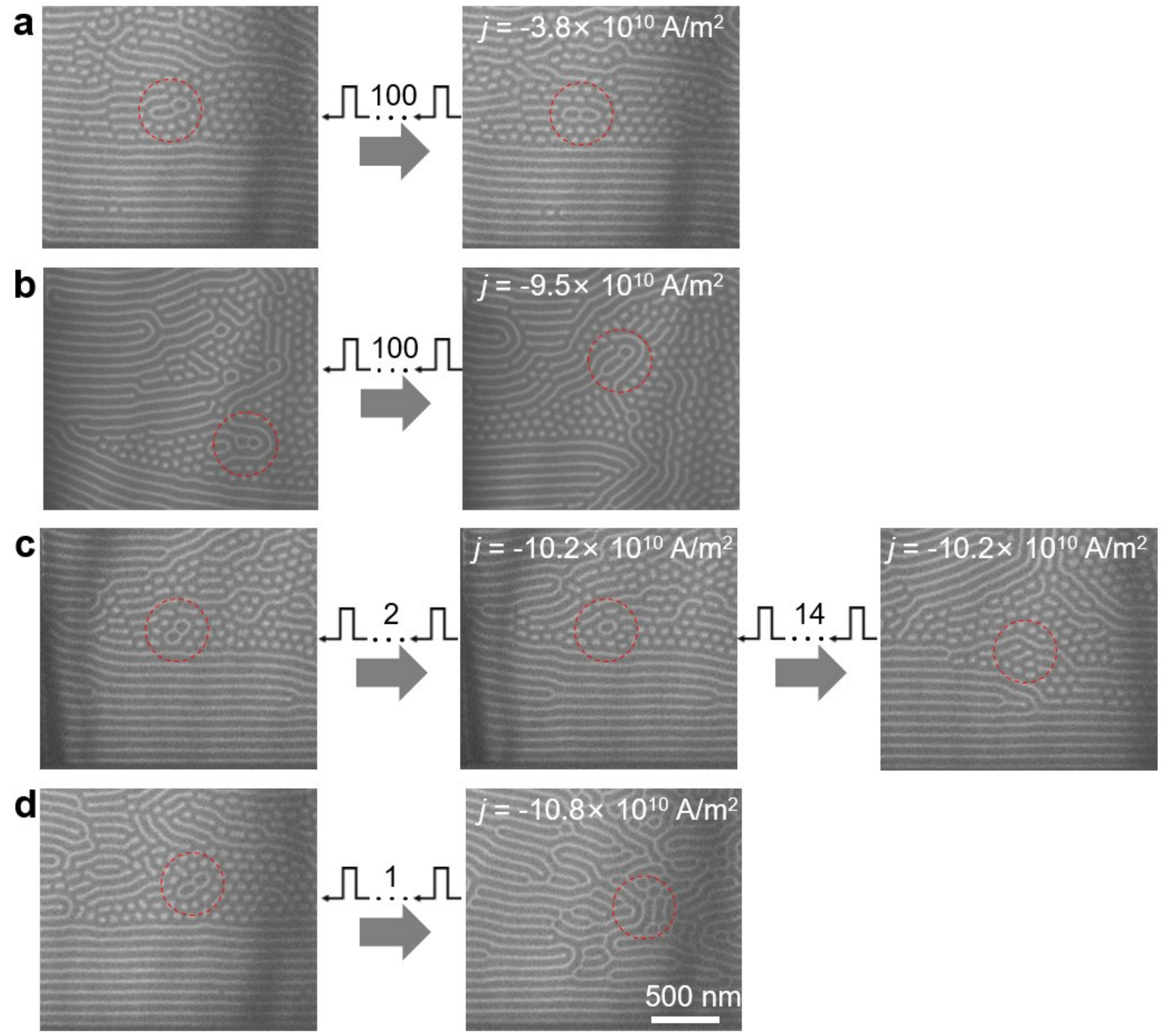

**Supplementary Fig. 19 | Effect of current density on the stability of the −*S*(2) skyrmion bag for a fixed pulse width of 20 ns.** The −*S*(2) skyrmion bag retains its topological state after applying 100 current pulses with densities of $-3.8\times 10^{10}$ A/m² (a) and $-9.5\times 10^{10}$ A/m² (b). At a current density of $-10.2\times 10^{10}$ A/m² (c), the bag first transforms into a −*S*(1) bag after 2 pulses and subsequently into a $Q = 1$ skyrmion after 14 pulses. At a higher current density of $-10.8\times 10^{10}$ A/m² (d), the state is reset after a single pulse.

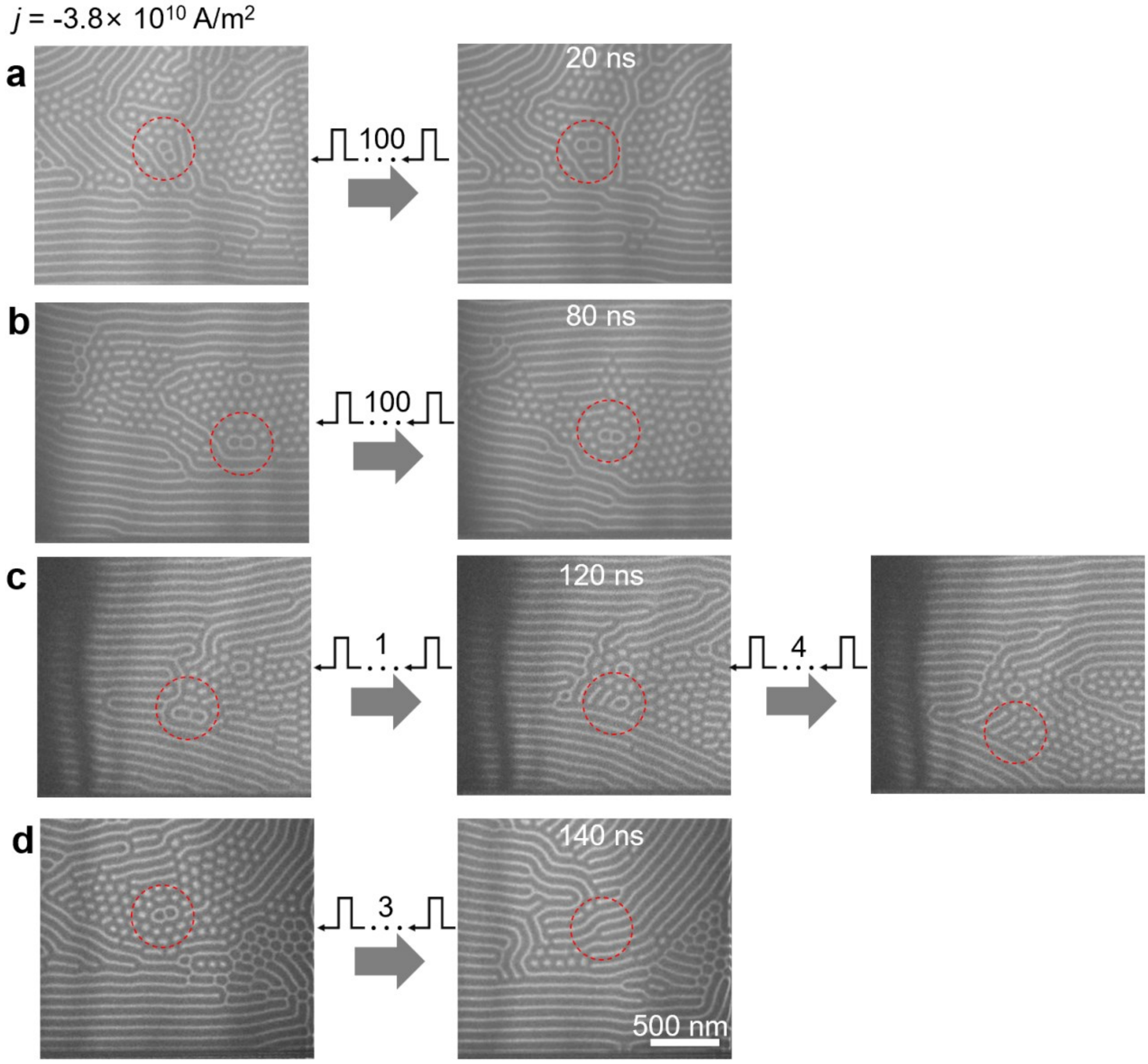

**Supplementary Fig. 20 | Effect of pulse width on the stability of the $-S(2)$ skyrmion bag for a fixed current density of $-3.8\times 10^{10}$ A/m$^2$.** The $-S(2)$ skyrmion bag retains its topological state after applying 100 current pulses with widths of 20 ns (a) and 80 ns (b). At a pulse width of 120 ns (c), the bag first transforms into a $-S(1)$ bag after 1 pulse and then into a $Q = 1$ skyrmion after 4 pulses. At a width of 140 ns (d), the state is reset after 3 pulses.

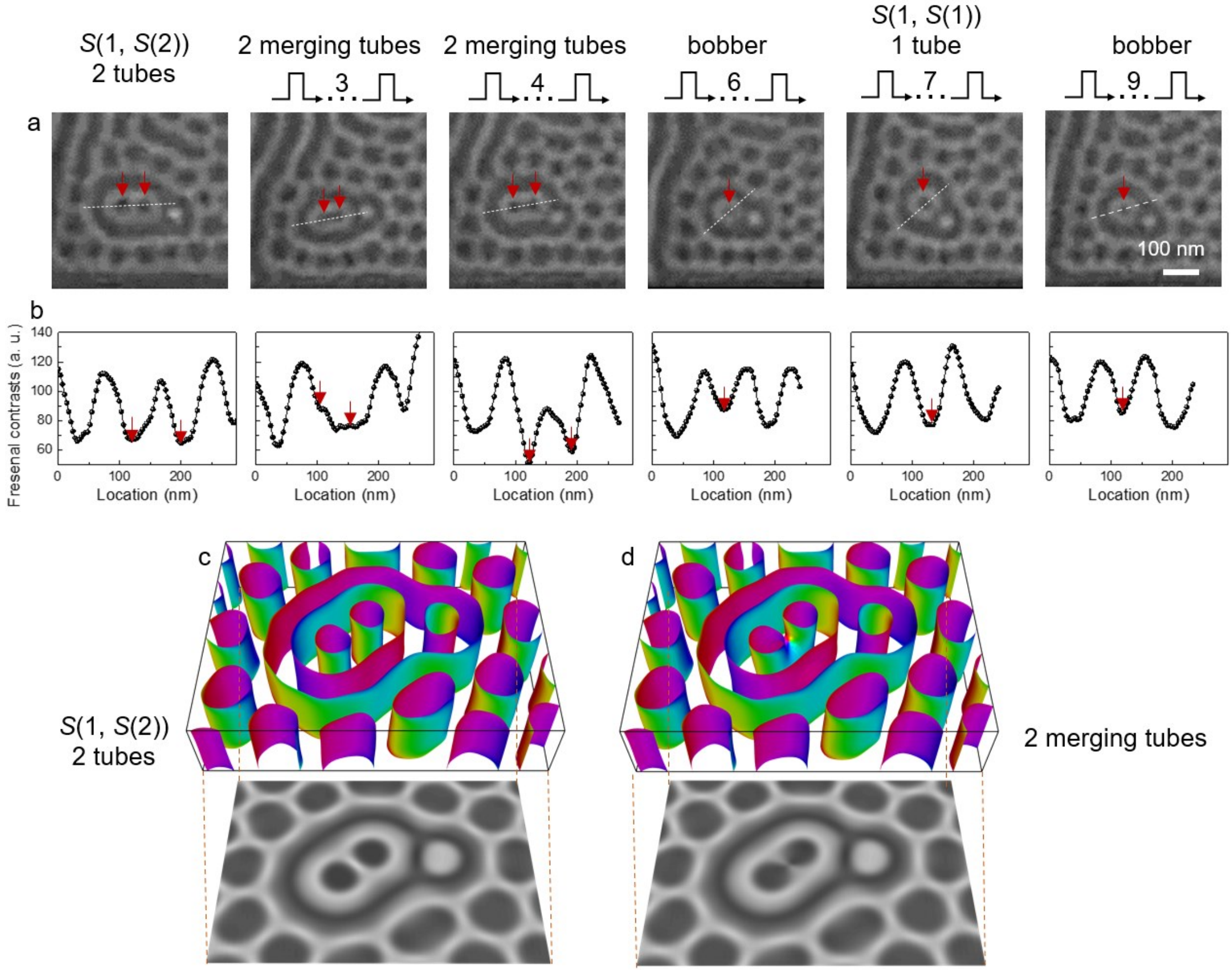

**Supplementary Fig. 21 | Current-induced collapse of a nested *S*(1, *S*(2)) skyrmion bag. a** Fresnel contrast during the collapse process. Defocused distance, −500 μm. **b** Line profiles of Fresnel contrast along the white dashed lines. The red arrows point out the location where the tubes are. **c, d** Simulated 3D magnetic iso-surface for $m_z$ = 0 and corresponding defocused Fresnel contrasts of *S*(1, *S*(2)) nested bags with two individual tubes (c) and two merging tubes (d).

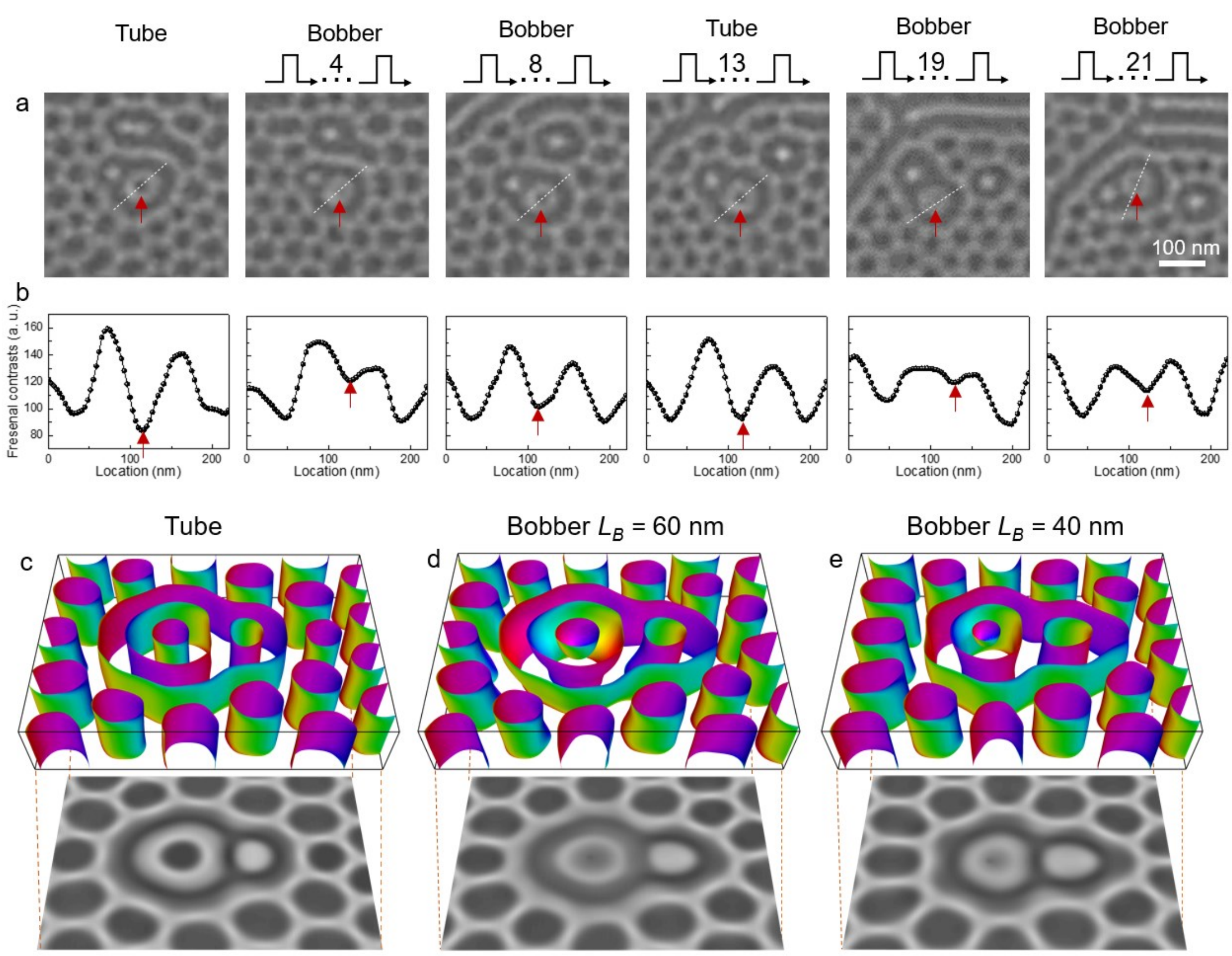

**Supplementary Fig. 22 | Current-induced collapse of a nested $S(1, S(1))$ skyrmion bag. a** Fresnel contrast during the collapse process. Defocused distance, −500 μm. **b** Line profiles of Fresnel contrast along the white dashed lines. The red arrows point out the location where the tubes/bobbers are. **c-e** Simulated 3D magnetic iso-surface for $m_z = 0$ and corresponding defocused Fresnel contrasts of $S(1, S(1))$ nested bags with 1 tube (c), 1 bobber with penetration depth $L_B = 56$ nm (d), and 1 bobber with penetration depth $L_B = 34$ nm (e).

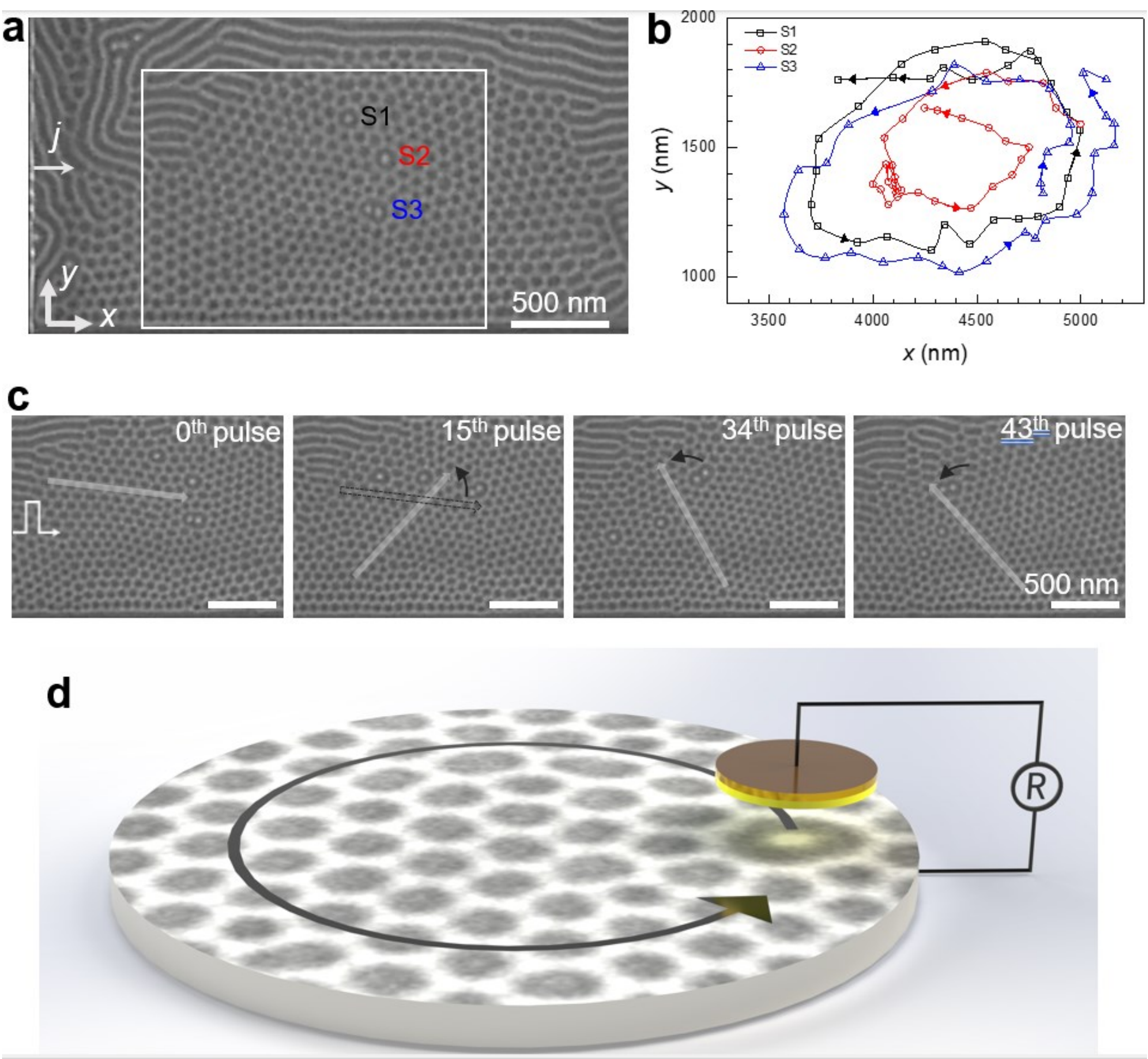

**Supplementary Fig. 23 | Current-driven rotation of skyrmion lattice and skyrmion bags at zero magnetic field. a** The overall view of the device. **b** The bag trajectories during the application of current pulses. S1, S2, and S3 represent the skyrmion bag as marked in (a). **c** Snapshots of the lattice and bag rotation driven by nanosecond current pulses. These Fresnel images were taken at a defocused distance of −500 μm. **d** Schematic device design based on skyrmion lattice rotation dynamics through the tunnel magnetoresistance detection of embedded skyrmion bags.

**Supplementary Movie captions:**

**Supplementary Movie V1 | Current-induced creation of *S*(1) and *S*(0, S(3)) magnetic skyrmion bags under zero magnetic field.** The process is initiated at 0:06 s with two high-amplitude current pulses ($5.5 \times 10^{10}$ A/m$^2$) to establish the initial magnetic state. Following this, a train of current pulses at a density of $4.5 \times 10^{10}$ A/m$^2$ drives the system dynamics (70 ns pulse duration, 1 Hz repetition rate).

**Supplementary Movie V2 | Current-induced creation of *S*(1) and *S*(0, *S*(4)) magnetic skyrmion bags.** The pulse duration is set to 70 ns with a frequency of 1 Hz, and the pulse amplitude is $4.5 \times 10^{10}$ A/m$^2$.

**Supplementary Movie V3 | Current-induced creation of *S*(1), *S*(2) and *S*(1, *S*(1)) magnetic skyrmion bags.** The pulse duration is set to 70 ns with a frequency of 1 Hz, and the pulse amplitude is $4.5 \times 10^{10}$ A/m$^2$.

**Supplementary Movie V4 | Current-induced dynamics on helix under the stimuli of high-density current.** The pulse duration is set to 70 ns, and the pulse amplitude is $5.6 \times 10^{10}$ A/m$^2$.

**Supplementary Movie V5 | Current-induced transformation from *S*(6) to *S*(5) bags.** The pulse duration is set to 70 ns with a frequency of 1 Hz, and the pulse amplitude is $4.5 \times 10^{10}$ A/m$^2$.

**Supplementary Movie V6 | Current-induced transformation from *S*(3) to *S*(2) bags.** The pulse duration is set to 70 ns with a frequency of 1 Hz, and the pulse amplitude is $4.5 \times 10^{10}$ A/m$^2$.

**Supplementary Movie V7 | Current-induced transformation from *S*(0, *S*(2)) to *S*(1) bags.** The pulse duration is set to 70 ns with a frequency of 1 Hz, and the pulse amplitude is $4.5 \times 10^{10}$ A/m$^2$.

**Supplementary Movie V8 | Current-induced transformation from *S*(1, *S*(2)) to *S*(1, *S*(1)) bags.** The pulse duration is set to 70 ns with a frequency of 1 Hz, and the pulse amplitude is $4.5 \times 10^{10}$ A/m$^2$.

**Supplementary Movie V9 | Current-induced dynamics of the *S*(1, *S*(1)) bag.** The pulse duration is set to 70 ns with a frequency of 1 Hz, and the pulse amplitude is $4.5 \times 10^{10}$ A/m$^2$.

**Supplementary References**